\documentclass[letterpaper,english,reprint, aps]{revtex4-1}
\usepackage[T1]{fontenc}
\usepackage[latin9]{inputenc}
\usepackage{amsmath}
\usepackage{amssymb}
\usepackage{graphicx}

\usepackage{hyperref,graphicx}
\hypersetup{
	colorlinks=true,
  linkcolor=blue,
  citecolor=blue,
  urlcolor=blue,
}



\usepackage{babel}
\begin{document}

\preprint{APS/123-QED}

\title{Anomalous diffusion in Davydov quantum molecular chain model}

\author{Sho Nakade$^{1,\ast}$, Kazuki Kanki$^{1}$, Satoshi Tanaka$^{1}$, Tomio Petrosky$^{2,3}$}

\affiliation{$^{1}$Department of Physical Science, Osaka Prefecture University,
Gakuen-cho 1-1, Sakai 599-8531, Japan\\
$^{2}$Center for Complex Quantum Systems,
The University of Texas at Austin, Austin, Texas 78712, USA\\
$^{3}$Institute of Industrial Science, The University of Tokyo,
5-1-5 Kasiwa 277-851, Japan}

\date{\today}
\begin{abstract}
We discuss
anomalous
relaxation processes in Davydov one-dimensional
chain molecule that consists of an exciton and
an acoustic phonon field as a thermal reservoir in the chain.
We derive a kinetic equation for the exciton using
the complex spectral representation of the Liouville-von Neumann
operator. Due to the one-dimensionality, the momentum
space separates into infinite sets of disjoint irreducible subspaces dynamically
independent of one another. Hence, momentum relaxation occurs only within each subspace
toward the Maxwell distribution.
We obtain a hydrodynamic mode with transport coefficients,
a sound velocity and a diffusion coefficient, defined in each
subspace. Moreover, because the sound velocity
has momentum dependence, phase mixing affects
the broadening of the spatial distribution of the exciton in addition to the diffusion process.
Due to the phase mixing the increase
rate of the mean-square displacement of the exciton increases linearly with time and diverges in the
long-time limit.
\end{abstract}



\maketitle

\section{Introduction}

The irreversible transport property in  one-dimensional (1D) systems has historically attracted many physicists because of their mathematical simplicity and some unique anomalies that appear due to extremely low dimensionality \cite{1953GKleinIPrigogine,1962IPrigogine,
1967JLLebowitzJKPercus,1978PResiboisMMareschal,
1997TPetroskyGOrdonez,2016KHashimotoKKankiSTanakaTPetrosky}.
For example, the collision operator in the kinetic equation disappears for a 1D gas that consists of particles of a same kind.
The reason is that the momentum distribution function cannot change in time, since the momenta of the particles are simply exchanged during the collision process in the 1D system.
However, this is not the case for quantum 1D systems, because there is the forward scattering in addition to the backward scattering in quantum mechanics~\cite{2016KHashimotoKKankiSTanakaTPetrosky}.
Hence, the irreversibility is purely a quantum effect for this case.

Other example is an anomalous diffusion process that has been recently pointed out by Pouthier  in a study of the transport process in the Davydov model \cite{2009VPouthier}.  This model has been introduced by Davydov for studying a mechanism of bioenergy transfer in a 1D protein molecular chain \cite{1977ASDavydov,1992ACScott}.  Pouthier has shown through a numerical calculation that
the diffusion process of the
vibronic exciton propagation in this 1D system is anomalous in the sense that the
phenomenological diffusion coefficient defined as the increase rate
of the exciton mean-square displacement as
\begin{equation}
  D^{(x)}(t)\equiv\frac{1}{2}\frac{d}{dt}\left\langle (X-\langle X\rangle_{t})^{2}\right\rangle _{t},
  \label{eq:phenomenological diffusion coefficient}
\end{equation}
increases linearly in time,
where $\langle\cdots\rangle_{t}$ denotes an average taken over the
time-depending density matrix. Hence the phenomenological diffusion coefficient diverges in the long-time limit.

In this paper we present a detailed theoretical analysis of this anomalous transport process. Our discussion is based on the complex spectral representation of the Liouville-von Neumann operator (Liouvillian) \cite{1997TPetroskyIPrigogine}.  This representation gives us a microscopic foundation of the irreversible
kinetic theory \cite{1997TPetroskyIPrigogine}. The complex spectral analysis of the Liouvillian shows that the collision operator in the
kinetic equation is just the effective Liouvillian defined by Eq. (\ref{eq:effective Liouvillian}), and hence, the spectrum of the collision operator coincides with that of the Liouvillian.

We will show that there exist well-defined transport coefficients including the diffusion coefficient and the hydrodynamic sound velocity that appear in the transport equation, in spite of the fact that the phenomenological diffusion coefficient~\eqref{eq:phenomenological diffusion coefficient} diverges in the long-time limit.
Indeed, we will show that the time dependence of $D^{(x)}(t)$ is given by
\begin{equation}
  D^{(x)}(t)=
  \bar{D}+t\bigl\langle (\sigma(P)-\bar{\sigma})^{2}\bigr\rangle_{\rm eq},
  \label{eq:phenomenological diffusion coefficient in 1D system}
\end{equation}
where
\begin{equation}
 \bar{D}\equiv \bigl\langle D(P)\bigr\rangle_{\rm eq},
 \\\ \
 \bar{\sigma}\equiv
 \bigl\langle\sigma(P)\bigr\rangle_{\rm eq},
\end{equation}
and $D(P)$ and $\sigma(P)$ denote the momentum dependent diffusion
coefficient and the hydrodynamic sound velocity,
respectively,
obtained from the complex spectrum  analysis of the Liouvillian. Here,
$\langle\cdots\rangle_{\rm eq}$ denotes an average taken over the equilibrium
state for the momentum distribution function.
The explicit forms of the hydrodynamic sound velocity
$\sigma(P)$ and the diffusion coefficient $D(P)$ are given in  Eqs. (\ref{eq:sound velocity})
and (\ref{eq:diffusion coefficient}).

In Eq.~\eqref{eq:phenomenological diffusion coefficient in 1D system} the momentum dependent sound velocity, which is a unique result for the 1D system,
is essential to understand the origin of the anomaly in the phenomenological diffusion coefficient. Indeed, it is well-known that in systems that have more than one dimension, the hydrodynamic sound mode appears in the spectrum of the irreversible collision operator in the kinetic equation when the collisional invariant has a degeneracy \cite{1977PResiboisMdeLeenery}. However, we found that the non-vanishing hydrodynamic sound velocity in our system is not due to a degeneracy. This is because of the one-dimensional restriction on the momenta contributing to the collision operator through the resonance condition.
For systems in more than one dimensions there is no such restriction in momentum space.
As a result, the momentum space in our 1D system is decomposed into infinitely many subset due to the resonance condition.
Then, this new mechanism in our 1D system leads to a unique hydrodynamic mode with a momentum dependent sound velocity.
This is not the case in systems in more than one dimensions.

As will be shown, the momentum-dependent sound velocity leads to the phase mixing during the time evolution of the distribution function. This then leads to  a spreading of the distribution function in space as a reversible process, in addition to the irreversible spreading due to the the diffusion process.
This phase mixing is the origin of the anomaly observed in the phenomenological diffusion coefficient (\ref{eq:phenomenological diffusion coefficient}).

In our previous works we have obtained the hydrodynamic mode of the effective Liouvillian associated to the inhomogeneity,
 and described the propagation of
hydrodynamic sound waves without taking account of diffusive relaxation
of the hydrodynamic mode \cite{2008STanakaKKankiTPetrosky,2009STanakaKKankiTPetrosky}.
In this paper we go beyond
the previous work by taking account of the diffusion process.

In this paper we restrict our interest to a situation where the exciton weakly
couples to the phonon.

The present paper is organized as follows:
in Sec.
\hyperlink{II}
{II}, the Davydov Hamiltonian is introduced.
Then, the short summary of the complex spectral representation of the Liouvillian is presented.
In Sec.
\hyperlink{III}
{III}, a kinetic equation for the spatially inhomogeneous distribution is investigated.
We will see that the resonance condition for the 1D system leads to the separation of the momentum space
into infinite sets of disjoint subspaces.
Moreover, we will show that the time evolution of the reduced density matrix of the exciton in the hydrodynamic regime
obeys a convection-diffusion equation with a sound velocity and
a diffusion coefficient.
In Sec.
\hyperlink{IV}
{IV}, the phenomenological diffusion coefficient defined by Eq. (\ref{eq:phenomenological diffusion coefficient})
is analyzed in terms of the transport coefficients in the kinetic equation
in the case the initial condition is given as a minimum uncertain wave packet.
Section
\hyperlink{V}
{V} is devoted to discussions about the relation between the
divergence of the phenomenological diffusion coefficient and
the one-dimensionality of the system.
There we will emphasize that the non-vanishing of the hydrodynamic sound velocity in this specific system is due to the disappearance of the momentum inversion symmetry in each subspace because of the one-dimensionality of the system.
%

In Appendix
\hyperlink{A}
{A}, we summarize the time evolution of the momentum distribution of the exciton.
In Appendix
\hyperlink{B}
{B}, a detailed derivation of the analytic expression of the phenomenological diffusion coefficient is presented when the initial condition is given as a minimum uncertain wave packet.
In Appendix
\hyperlink{C}
{C}, we extend the result obtained in Appendix B to more general initial conditions with arbitrary square integrable functions.

\hypertarget{II}{}
\section{Davydov Model and complex spectral analysis of the Liouvillian}

\subsection{A. Davydov model}

In this paper, we consider the Davydov Hamiltonian that is a simple
model for a molecular chain \cite{1990PLChristiansenACScott,1992ACScott}.
We assume that
an exciton weakly interacts with phonons of an underlying
lattice. We consider relaxation dynamics of the exciton on this one-dimensional
quantum molecular chain. A Hamiltonian of
the
%
system is given by \cite{1990PLChristiansenACScott,1992ACScott,2009STanakaKKankiTPetrosky}
\begin{equation}
 H=H_{0}+gV,\label{eq:Total Hamiltonian}
\end{equation}
\begin{align}
 H_{0} & =\sum_{p}\varepsilon_{p}|p\rangle\langle p|
         +\sum_{q}\hbar\omega_{q}a_{q}^{\dagger}a_{q},\\
 V & =\sqrt{\frac{2\pi}{L}}\sum_{p,q}V_{q}|p+\hbar q\rangle\langle p|
      (a_{q}+a_{-q}^{\dagger}).
\end{align}
where $g$ is a dimensionless coupling parameter which is introduced
to indicate the order of the interaction between the exciton and phonon
of underlying lattice for the convenience of the perturbation analysis.
After finishing the weak-coupling approximation, we set $g=1$. The
notation $L$ denotes the length of the chain. The momentum of the
exciton is designated by $p$ and its state by $|p\rangle$, and $a_{q}$
and $a_{q}^{\dagger}$ are annihilation and creation operators of
the phonon with wave vector $q$.
We assume that state $|p\rangle$ is normalized by the Kronecker delta.

We consider the case where dispersion
relations are given by
\begin{equation}
 \varepsilon_{p}=\frac{p^2}{2m},\ \ \omega_{q}=c|q|,
\end{equation}
 with the effective mass $m$ for the exciton, and linear dispersion
for acoustic phonon with the speed of sound $c$. We assume a deformation
potential type \cite{1993GeraldDMahan} for the coupling between the
exciton and the phonons as
\begin{equation}
V_{q}\equiv\Delta_{0}|q|\sqrt{\frac{\hbar}{4\pi\rho_{M}\omega_q}},
\end{equation}
where $\Delta_{0}$ is the coupling constant, and $\rho_{M}$ is the
molecular mass density, i.e mass per unit length, of the lattice.

Let us introduce the following units: time unit $\hbar/(mc^{2})$,
length unit $\hbar/(mc)$, momentum unit $mc$, energy unit $mc^{2}$,
and temperature unit $mc^{2}/k_{\rm B}$
with the Boltzmann constant $ k_{\rm B}$.
With these units, $m,c,\hbar$, and $k_{{\rm B}}$ are dimensionless and that correspond to the choice of $m=1,c=1,\hbar=1,\ {\rm and}\ k_{{\rm B}}=1$.

According to Refs. \cite{1992ACScott,1990PLChristiansenACScott},
the energy unit can be evaluated as $mc^{2}\simeq3.0\times10^{-21}{\rm J}$.
Thus, the temperature unit can be estimated as $mc^{2}/k_{\rm B}\simeq220{\rm K}$.
With this unit, the physiological temperature $T=310{\rm K}$ is around
$k_{\rm B}T/mc^{2}\simeq1.41$. Therefore we focus our attention
on the temperature domain $1\lesssim k_{\rm B}T/mc^{2}\lesssim2$
 corresponds to  $220{\rm K}\lesssim T \lesssim 440{\rm K}$
 as the
  temperature domain for this model.

We impose a periodic boundary condition with period $L$ leading to
discrete momenta and wave numbers $p/\hbar$, $q=2\pi j/L$ with $j=0,\pm1,\pm2,\cdots$.
We consider a long chain and will approximate the length as $L\rightarrow\infty$,
then we will replace a summation over a momentum and a wave number
with an integration at an appropriate stage.
\begin{align}
 \frac{2\pi\hbar}{L}\sum_{P} & \rightarrow\int dP,\ \ \frac{L}{2\pi\hbar}\delta_{P,P^{\prime}}^{\rm Kr}
  \rightarrow\delta(P-P^{\prime}),\\
 \frac{2\pi}{L}\sum_{q} & \rightarrow\int dq,  \ \ \ \frac{L}{2\pi}\delta_{q,q'}^{\rm Kr}
  \rightarrow\delta(q-q'),
\end{align}
where $\delta_{q,q'}^{\rm Kr}$ denotes the Kronecker delta, and $\delta(q-q')$ is the delta function.

The time evolution of the density operator $\rho(t)$ of the total
system obeys the quantum Liouville equation
\begin{equation}
 i\frac{\partial}{\partial t}\rho(t)={\cal L}_{H}\rho(t),
 \label{eq:quantum Liouville equation}
\end{equation}
where the Liouvillian ${\cal L}_{H}$ is defined by
the commutation with the Hamiltonian as
\begin{equation}
 {\cal L}_{H}\rho(t)\equiv\frac{1}{\hbar}[H,\rho(t)].
\end{equation}
Corresponding to the decomposition in Eq. (\ref{eq:Total Hamiltonian}),
Liouvillian is decomposed as
\begin{equation}
 {\cal L}_{H}={\cal L}_{0}+g{\cal L}_{V},
\end{equation}
where ${\cal L}_{0}\equiv{\cal L}_{H_0}$.

We focus our attention on the time evolution of the reduced density
operator for the exciton, which is defined by
\begin{equation}
 f(t)\equiv{\rm Tr}_{\rm ph}[\rho(t)],
\end{equation}
where ${\rm Tr}_{\rm ph}$ indicates that the trace is taken over
all the phonon modes. The phonons are assumed to be in thermal equilibrium
in the initial state represented by
\begin{equation}
 \rho_{{\rm ph}}^{{\rm eq}}=
  \frac{1}{Z_{{\rm ph}}}\exp
  \left[
   -\sum_{q}\hbar\omega_{q}b_{q}^{\dagger}b_{q}/k_{{\rm B}}T
  \right],
 \label{eq:phonon equibrium density matrix}
\end{equation}
with the partition function
\begin{equation}
 Z_{{\rm ph}}\equiv\prod_{q}(1-\exp[-\hbar\omega_{q}/k_{{\rm B}}T])^{-1}.
\end{equation}

One can readily show that the deviation of the
phonon distribution from the thermal equilibrium is proportional to
$1/L$ during the time evolution, so that the phonon distribution
remains in the thermal equilibrium in the limit $L\rightarrow\infty$.

We follow the time evolution of the reduced density operator in terms
of a vector representation $|f(t)\rangle\!\rangle$ in the Liouville
space. In this space, the inner product of the linear
operators
$A$ and $B$
in the wave function space
is defined by
\begin{equation}
 \langle\!\langle A|B\rangle\!\rangle={\rm Tr}[A^{\dagger}B].
\end{equation}

We express the reduced density operator in terms of the Wigner representation
in the momentum space of the exciton \cite{1997TPetroskyIPrigogine,1962IPrigogine} as
\begin{equation}
 f_{k}(P,t)\equiv\langle\!\langle k,P|f(t)\rangle\!\rangle
           \equiv\left( P+\frac{\hbar k}{2}\right|f(t)\left|P-\frac{\hbar k}{2}\right),
           \label{eq:Wigner expression}
\end{equation}
where the Wigner basis is defined as
\begin{equation}
 |k,P\rangle\!\rangle\equiv|P+\hbar k/2)( P-\hbar k/2|.
\end{equation}
The round bracket is defined as
\begin{equation}
 |p)\equiv\sqrt{\frac{L}{2\pi\hbar}}|p\rangle,
\end{equation}
and normalized by the delta function in the limit $L\to \infty $ as
\begin{equation}
 (p|p^{\prime})=\delta(p-p^{\prime}).
\end{equation}
In this representation, the component $f_{0}(P,t)$ is a momentum
distribution of the exciton.

The Fourier transform
\begin{equation}
 f^{W}(X,P,t)\equiv\frac{1}{2\pi}
 \int_{-\infty}^{\infty}f_{k}(P,t)\exp[ikX]dk,
 \label{eq:fourier transform to Wigner distribution function}
\end{equation}
is the Wigner distribution function in the ``quantum phase space'' $(X, P)$, which corresponds to the distribution function in the classical phase space \cite{1997TPetroskyIPrigogine}. 

\subsection{B. Complex spectral analysis}

The kinetic theory in non-equilibrium statistical physics is discussed
in terms of the complex spectral representation of the Liouville-von
Neumann operator (Liouvillian). The
eigenvalue problem of
the Liouvillian for each correlation subspace $\mu$ is given by
\begin{equation}
 {\cal L}_H|F_{j}^{(\mu)}\rangle\!\rangle=
 z_{j}^{(\mu)}|F_{j}^{(\mu)}\rangle\!\rangle,\ \ \ \langle\!\langle\tilde{F_j}^{(\mu)}|{\cal L}_{H}=
 \langle\!\langle\tilde{F_j}^{(\mu)}|z_{j}^{(\mu)},
 \label{eq:eigenvalue problem of total liouvillian}
\end{equation}
 where the indices $\mu$ and $j$ specify the eigenvalue and eigenstate.
%
The Liouvillian may have generally complex eigenvalues ${\rm Im}z_{j}^{(\mu)}\neq0$.
Hence the left eigenstate is not a hermitian conjugate of the right
eigenstate. The right and left eigenstates, $|F_{j}^{(\mu)}\rangle\!\rangle$
and $\langle\!\langle\tilde{F_{j}}^{(\mu)}|$, are biorthonormal sets
satisfying
\begin{equation}
 \langle\!\langle
  \tilde{F_j}^{(\mu)}|F_{j^{\prime}}^{(\mu^{\prime})}
 \rangle\!\rangle=
 \delta_{j,j^{\prime}}^{{\rm Kr}}\delta_{\mu,\mu^{\prime}}^{{\rm Kr}},\ \ \ \sum_{\mu,j}|F_{j}^{(\mu)}\rangle\!\rangle
 \langle\!\langle\tilde{F_{j}}^{(\mu)}|=1.
 \label{eq:bi-orthoganality and completeness relations of F}
\end{equation}

 The eigenvalue problem in Eq. (\ref{eq:eigenvalue problem of total liouvillian})
can be deformed by Brillouin-Wigner-Feshbach projection operator method of the effective Liouvillian \cite{1997TPetroskyIPrigogine,2010TPetrosky,2009STanakaKKankiTPetrosky}. For this, we introduce the projection operator ${\cal P}^{(\mu)}$ and its compliment projection operator ${\cal Q}^{(\mu)}$ associated to the eigenstates of ${\cal L}_{0}$, as
\begin{equation}
 {\cal P}^{(\mu)} + {\cal Q}^{(\mu)}= 1,
\end{equation}
which  satisfy the following relations
\begin{align}
 {\cal L}_{0}{\cal P}^{(\mu)} & =
  {\cal P}^{(\mu)}{\cal L}_{0}=w_{\mu}{\cal P}^{(\mu)},\\
 {\cal P}^{(\mu)}{\cal P}^{(\mu^{\prime})} & =
  {\cal P}^{(\mu)}\delta_{\mu,\mu^{\prime}}^{\rm Kr},\\
 {\cal Q}^{(\mu)}{\cal Q}^{(\mu^{\prime})} & =
  {\cal Q}^{(\mu)}\delta_{\mu,\mu^{\prime}}^{\rm Kr},\\
 {\cal P}^{(\mu)}{\cal Q}^{(\mu)} & ={\cal Q}^{(\mu)}{\cal P}^{(\mu)}=0,
\end{align}
 where $w_{\mu}$ is an eigenvalue of unperturbed Liouvillian ${\cal L}_{0}$.

 Then, for example, operating ${\cal P}^{(\mu)}$ and ${\cal Q}^{(\mu)}$
 respectively from the left-hand side of the right eigenvalue equation in Eq. (\ref{eq:eigenvalue problem of total liouvillian}), we get a set of equations for ${\cal P}^{(\mu)}|F_{j}^{(\mu)}\rangle\!\rangle$ and  ${\cal Q}^{(\mu)}|F_{j}^{(\mu)}\rangle\!\rangle$.
 One can solve this set of equation in a simple algebra, and obtains an equation
\begin{equation}
 \Psi^{(\mu)}(z_{j}^{(\mu)})|\phi_{j}^{(\mu)}\rangle\!\rangle=
 z_{j}^{(\mu)}|\phi_{j}^{(\mu)}\rangle\!\rangle,
 \label{eq:eigenvalue problem of the effective Liouvillian}
\end{equation}
where
$|\phi_{j}^{(\mu)}\rangle\!\rangle \equiv
 \sqrt{N_{j}^{(\mu)}} {\cal P}^{(\mu)}|F_{j}^{(\mu)}\rangle\!\rangle$
with a suitable normalization constant $N_{j}^{(\mu)}$,
and $\Psi^{(\mu)}(z)$ is
the effective Liouvillian defined as
\begin{equation}
 \Psi^{(\mu)}(z)\equiv
 {\cal P}^{(\mu)}{\cal L}_{H}{\cal P}^{(\mu)}+
 {\cal P}^{(\mu)}g{\cal L}_{V}{\cal Q}^{(\mu)}{\cal C}^{(\mu)}(z){\cal P}^{(\mu)},
 \label{eq:effective Liouvillian}
\end{equation}
 with the creation-of-correlation operator \cite{1997TPetroskyIPrigogine}
defined as
\begin{equation}
 {\cal C}^{(\mu)}(z)\equiv
 \frac{1}{z-{\cal Q}^{(\mu)}{\cal L}_{H}{\cal Q}^{(\mu)}}
 {\cal Q}^{(\mu)}g{\cal L}_{V}{\cal P}^{(\mu)}.
 \label{eq:creation-of-correlation operator}
\end{equation}

The second term of the effective Liouvillian $\Psi^{(\mu)}(z_{j}^{(\mu)})$ in Eq. (\ref{eq:effective Liouvillian}) in the Liouville space corresponds to the self-energy part of the Hamilton in the wave function space. The effective Liouvillian is also known as the ``collision operator'' which is the central object in the non-equilibrium statistical mechanics.

By solving the set of the right eigenstate also for the ${\cal Q}^{(\mu)}$ component, and combining it with the ${\cal P}^{(\mu)}$ component, we obtain the right eigenstates of the total Liouvillian ${\cal L}_{H}$ as
\begin{equation}
 |F_{j}^{(\mu)}\rangle\!\rangle=
 \sqrt{N_{j}^{(\mu)}}
 \left({\cal P}^{(\mu)}+{\cal C}^{(\mu)}(z_{j}^{(\mu)})\right)
 |\phi_{j}^{(\mu)}\rangle\!\rangle.
\end{equation}
 The left eigenstates
of the total Liouvillian ${\cal L}_{H}$ can be obtain in the same
manner.

For our specific model, we have $(\mu)=(k,\nu)$, and the projection operator ${\cal P}^{(k,\nu)}$
defined as
\begin{equation}
 {\cal P}^{(k,\nu)}\equiv
 \sum_{N}\int dP|k,P\rangle\!\rangle\langle\!\langle k,P|\otimes
 |\{\nu\},\{N\}\rangle\!\rangle\langle\!\langle\{\nu\},\{N\}|,
\end{equation}
 where $|\{\nu\},\{N\}\rangle\!\rangle$ is the orthonormal basis
in the Liouville space for the phonon state defined as
\begin{equation}
 |\{\nu\},\{N\}\rangle\!\rangle\equiv
 |\{n\};\{n^{\prime}\}\rangle\!\rangle=
 |\{N+\frac{\nu}{2}\};\{N-\frac{\nu}{2}\}\rangle\!\rangle.
 \label{eq:wigner basis for phonon states}
\end{equation}
The notation $\{\cdots\}$ in Eq. (\ref{eq:wigner basis for phonon states})
denotes a direct product of all $q$ modes of phonon states.

Moreover, the eigenvalue of the unperturbed Liouvillian is given by
\begin{equation}
  w_{\mu}\equiv w_{k,P}+\nu\omega,
\end{equation}
with
\begin{align}
  w_{k,P} & \equiv
  \frac{1}{\hbar}(\varepsilon_{P+\frac{\hbar k}{2}}-\varepsilon_{P-\frac{\hbar k}{2}})=k\frac{P}{m},\\
  \nu\omega & \equiv
  \sum_{q}\nu_{q}\omega_{q}.
\end{align}

Using these eigenstates of the Liouvillian, one can construct the irreversible kinetic equation for the non-equilibrium system.
Indeed, by using the completeness relation of Eq. (\ref{eq:bi-orthoganality and completeness relations of F})
and the fact that the eigenvalues of the effective Liouvillian are
equivalent to the eigenvalues of the total Liouvillian, one can derive a kinetic equation for the ${\cal P}^{(\mu)}$ of the density matrix through the complex eigenvalue problem of the Liouvillian \cite{1997TPetroskyIPrigogine,2010TPetrosky,2009STanakaKKankiTPetrosky}. Then one can see that the effective Liouvillian reduces to the collision operator in the kinetic equation.

For this derivation, we note that
the eigenvalue problem of the collision operator (\ref{eq:eigenvalue problem of the effective Liouvillian})
is nonlinear problem in the sense that the eigenvalue $z_{j}^{(\mu)}$
appears in the collision operator.

However, in the weak-coupling case,
a linear approximation of the eigenvalue problem of the collision
operator can be taken because the eigenvalue in the collision operator
can be approximated by the eigenvalue of the unperturbed Liouvillian
${\cal L}_{0}$. For this case, we also may expand $\Psi^{(\mu)}(z)$
in Eq. (\ref{eq:effective Liouvillian}) into a power series in the coupling
parameter $g$
\cite{1954LVanHove,1962IPrigogine,
1992CCohenTannoudjiJDupontRocGGrynberg,
1997TPetroskyIPrigogine,2010TPetrosky}.
Then, we obtain up to the second order in $g$ as
\begin{widetext}
\begin{align}
 \Psi^{(\mu)}(z_{j}^{(\mu)}) & \simeq
 \Psi_{2}^{(\mu)}(w_{\mu}+i0^{+})\nonumber\\
 & =
 {\cal P}^{(\mu)}{\cal L}_{0}{\cal P}^{(\mu)}+
 g^{2}{\cal P}^{(\mu)}{\cal L}_{V}{\cal Q}^{(\mu)}
  \frac{1}{w_{\mu}+i0^{+}-{\cal L}_{0}}
  {\cal Q}^{(\mu)}{\cal L}_{V}{\cal P}^{(\mu)}.
 \label{eq:linear approximated effective Liouvillian evaluated up to the second-order of the interaction}
\end{align}
\end{widetext}
Consequently, Markov kinetic equation in the weak coupling approximation
is derived as
\begin{equation}
 i\frac{\partial}{\partial t}{\cal P}^{(\mu)}|\rho(t)\rangle\!\rangle=
 \Psi_{2}^{(\mu)}(w_{\mu}+i0^{+})
  {\cal P}^{(\mu)}|\rho(t)\rangle\!\rangle.
 \label{eq:Markov kinetic equation in the weak coupling approximation}
\end{equation}
 The detailed derivation of the kinetic
equation can be found in Refs. \cite{1997TPetroskyIPrigogine,2010TPetrosky,2009STanakaKKankiTPetrosky}.

By taking the partial trace on the equilibrium phonon distribution, we obtain the kinetic equation
for the reduced distribution ${\cal P}^{(k,0)}|f(t)\rangle\!\rangle$ for $\mu = (k,0)$, which is given by
\begin{equation}
 i\frac{\partial}{\partial t}{\cal P}^{(k,0)}|f(t)\rangle\!\rangle=
  \overline{\Psi}_{2}^{(k)}(w_{k,P}+i0^{+})
   {\cal P}^{(k,0)}|f(t)\rangle\!\rangle,
 \label{eq:for fk}
\end{equation}
where
\begin{widetext}
\begin{align}
 \overline{\Psi}_{2}^{(k)}(w_{k,P}+i0^{+}) & \equiv
  {\rm Tr}_{\rm ph}\left[\Psi_{2}^{(k,0)}(w_{k,P}+i0^{+})\right]\nonumber\\
 & =
  {\rm Tr}_{\rm ph}
  \left[
   w_{k,P}{\cal P}^{(k,0)}+
   g^{2}{\cal P}^{(k,0)}{\cal L}_{V}{\cal Q}^{(k,0)}
    \frac{1}{w_{k,P}+i0^{+}-{\cal L}_0}{\cal Q}^{(k,0)}{\cal L}_{V}{\cal P}^{(k,0)}\rho_{\rm ph}^{\rm eq}
  \right],
 \label{eq:collison operator}
\end{align}
 with the phonon equilibrium distribution $\rho_{{\rm ph}}^{{\rm eq}}$
given by Eq. (\ref{eq:phonon equibrium density matrix}).

By multiplying $\langle\!\langle k,P|$ from the left of Eq. (\ref{eq:for fk}), we have
\begin{equation}
 i\frac{\partial}{\partial t}f_{k}(P,t)={\cal K}_P^{(k)}f_{k}(P,t),
 \label{eq:kneq0 kinetic eq}
\end{equation}
 where ${\cal K}_P^{(k)}$ is a matrix element of the collision operator
(\ref{eq:collison operator}), and defined as
\begin{align}
 {\cal K}_P^{(k)}\delta(P-P^{\prime})& \equiv
  \langle\!\langle
   k,P|\overline{\Psi}_{2}^{(k)}(w_{k,P}+i0^{+})|k,P^{\prime}
  \rangle\!\rangle   \nonumber \\
 & =
  w_{k,P}\delta(P-P^{\prime})+
  g^{2}
   \langle\!\langle
    k,P|{\rm Tr}_{{\rm ph}}
     \left[
       {\cal P}^{(k,0)}{\cal L}_{V}{\cal Q}^{(k,0)}
       \frac{1}{w_{k,P}+i0^{+}-{\cal L}_{0}}{\cal Q}^{(k,0)}
       {\cal L}_{V}{\cal P}^{(k,0)}\rho_{{\rm ph}}^{{\rm eq}}
     \right]
    |k,P^{\prime}
   \rangle\!\rangle.
 \label{eq:kneq0 collision op}
\end{align}
\end{widetext}
The first term of Eq. (\ref{eq:kneq0 collision op}) is the flow term,
which comes from the unperturbed Liouvillian ${\cal L}_{0}$.
The second term of Eq. (\ref{eq:kneq0 collision op}) is the collision term,
which comes from the interaction part in the Liouvillian ${\cal L}_{V}$.


\hypertarget{III}{}
\section{Exciton propagation in the hydrodynamic regime}

\subsection{A. Transport coefficients in kinetic equation}

In this section, we consider the time evolution of spatially inhomogeneous
distribution function $f_{k}(P,t)$.
The kinetic equation (\ref{eq:kneq0 kinetic eq})
governing the time evolution of $f_{k}(P,t)$ has the form of the
Boltzmann equation that consists of the flow term and the collision term.
The flow term is time-symmetric, while the collision term breaks time-reversal
symmetry. We treat the flow term as a perturbation to the collision
term in the hydrodynamic regime where the length scale $L_{{\rm h}}$ of the spatial
inhomogeneity is much longer than the mean-free path $L_{{\rm rel}}$ of the exciton,
\begin{equation}
  L_{{\rm h}}\gg L_{{\rm rel}}
\end{equation}
 where
\begin{equation}
  L_{{\rm rel}}\equiv\langle v\rangle\tau_{{\rm rel}},
  \label{eq:range of homogeniety}
\end{equation}
 with the average velocity $\langle v\rangle$ and the relaxation
time $\tau_{{\rm rel}}$ of the exciton defined in Eq. (\ref{eq:relaxation time}).

As shown in Appendix A, the spectrum (\ref{eq:spectrum of collision op for homogeneous system})
of the collision operator for the homogeneous system with $k=0$ is discrete.
This implies that the time scale of the
relaxation of the momentum distribution to the equilibrium is much
shorter than the time scale of relaxation of the spatial distribution
to the homogeneous distribution in the hydrodynamic regime.
As a result, the local equilibrium
is established before any appreciable change in spatial distribution
occurs.

For this hydrodynamic regime, one can treat the flow term as a perturbation to the collision term.
Then, we have calculated in our previous paper \cite{2009STanakaKKankiTPetrosky}
up to the first-order of the perturbation expansion with respect to the flow term,
and obtained a kinetic equation which has the form of a macroscopic linear wave equation
\begin{equation}
  \frac{\partial^{2}}{\partial t^{2}}f^{W}(X,P,t)=
  \sigma^{2}(P)\frac{\partial^{2}}{\partial X^{2}}f^{W}(X,P,t),
  \label{eq:macro linear wave eq}
\end{equation}
where $\sigma(P)$ is a hydrodynamic sound velocity.
It should be emphasized that the wave equation (\ref{eq:macro linear wave eq})
is for the probability distribution function,
and not for the wave amplitude, in spite of the fact that we are dealing with a quantum system.
This amazing feature is a direct consequence of the dissipative effect.

Note that the sound velocity depends on the momentum $P$.
This remarkable property is a characteristic of the one-dimensionality of the system.
Due to this momentum dependency, we have shown in Ref. \cite{2009STanakaKKankiTPetrosky}
that a phase mixing occurs in the exciton propagation similarly to the nonlinear dynamical system.

We now extend our previous result up to the second-order of the perturbation expansion with respect to the flow term
in the kinetic equation  (\ref{eq:kneq0 kinetic eq}). Then we will derive the convection-diffusion equation (\ref{eq:convection-diffusion equation}).


We apply the usual treatment in the hydrodynamic regime, i.e. we approximate
the collision operator for the $k$-component of the Wigner distribution function
as
\begin{widetext}
\begin{align}
  {\cal K}_P^{(k)}f_{k}(P,t)&=
   \int dP^{\prime}{\cal K}_P^{(k)}
    \delta(P-P^{\prime})f_{k}(P^{\prime},t) \nonumber\\
  &\approx
   \int dP^{\prime}w_{k,P}\delta(P-P^{\prime})f_{k}(P^{\prime},t)
   +\int dP^{\prime}{\cal K}_P^{(0)}
     \delta(P-P^{\prime})f_{k}(P^{\prime},t) \nonumber\\
  &=
   \left(k\frac{P}{m}+{\cal K}_P^{(0)}\right) f_{k}(P,t),
  \label{eq:hydrodynamic approximation of collision op}
\end{align}
where the collision term ${\cal K}_P^{(0)}$ is a difference operator, and acts on the momentum function as
\begin{align}
  {\cal K}_P^{(0)}f_{k}(P,t) & =
   -g^{2}\frac{2\pi i}{\hbar^{2}}\int\!\!dq|V_{q}|^{2}
    \left\{
     \delta\left(\frac{\varepsilon_{P}-\varepsilon_{P+\hbar   q}}{\hbar}+\omega_{q}\right)n(q)+
     \delta\left(\frac{\varepsilon_{P-\hbar q}-\varepsilon_{P}}{\hbar}+\omega_{q}\right)[n(q)+1]
    \right\}
   f_{k}(P,t)\ \nonumber \\
  & \ \ \
   +g^{2}\frac{2\pi i}{\hbar^{2}}\int\!\!dq|V_{q}|^{2}
    \left\{
     \delta\left(\frac{\varepsilon_{P-\hbar q}-\varepsilon_{P}}{\hbar}+\omega_{q}\right)n(q)f_{k}(P-\hbar q,t)\right.\nonumber \\
  & \ \ \ \ \ \ \ \ \ \ \ \ \ \ \ \ \ \ \ \ \ \ \ \ \ \ \
    \left.
     +\delta\left(\frac{\varepsilon_{P}-\varepsilon_{P+\hbar q}}{\hbar}+\omega_{q}\right)[n(q)+1]f_{k}(P+\hbar q,t)
    \right\} .
  \label{eq:concrete form of the collision op}
\end{align}
\end{widetext}

In Eq. (\ref{eq:concrete form of the collision op}), $n(q)$ denotes
the average number of phonons with a wave number $q$, and obeys the
Bose-Einstein distribution,
\begin{equation}
 n(q)\equiv\frac{1}{\exp[\hbar\omega_{q}/k_{{\rm B}}T]-1}.
\end{equation}
We note that, in the classical limit $\hbar\rightarrow0$, Eq. (\ref{eq:concrete form of the collision op})
becomes the form of $\int\!dq\ q\delta(q)$ and vanishes. Hence, the
dissipation for the weakly coupled system in 1D is purely a quantum
effect.

With the approximation (\ref{eq:hydrodynamic approximation of collision op}),
the eigenvalue problem for the collision operator ${\cal K}_P^{(k)}$ reduces to the eigenvalue problem,
\begin{equation}
\left[{\cal K}_P^{(0)}+k\frac{P}{m}\right]\phi_{j}^{(k)}(P)=z_{j}^{(k)}\phi_{j}^{(k)}(P).
\end{equation}

According to the resonance condition represented by the delta-functions
in the collision operator (\ref{eq:concrete form of the collision op}),
the state of the exciton can change from a state with a momentum $P$
only to two other states with momentum $-P\pm2mc$ via absorption
or emission of a phonon. This is a characteristic result due to one-dimensionality.
In two or more dimensional system, the resonance condition in the
collision operator connects all momentum states due to the angular degrees of freedom.

Starting with a momentum $P_{0}$, all the momenta which are coupled
successively through the collision operator are enumerated by applying
the following recursive formula:
\begin{equation}
  P_{\nu\pm1}=-P_{\nu}\pm(-1)^{\nu}2mc,
  \label{eq:1D discrete momenta}
\end{equation}
where $\nu$ is any integer. The solution of this recursive formula
with an initial value $P_{0}$ is given by
\begin{equation}
  P_{\nu}=(-1)^{\nu}(P_{0}-2\nu mc)\ \ \ \ \ \ \ \ (\nu=0,\pm1,\pm2,\cdots).
  \label{eq:discrete momentum states}
\end{equation}

A different choice of $P_{0}$ in the range $-mc\leq P_{0}\leq mc$
gives a different and disjoint set of momenta. Hence, due to the one-dimensionality
of the system, momentum space is separated into an infinite number
of disjoint subspaces. The components of the
distribution
function $f_{k}(P,t)$ with momenta $P$ in the set (\ref{eq:discrete momentum states})
connected to a single $P_{0}$ are independent of other components
with momenta connected to any other $P_{0}$. The full momentum dependence
can be obtained by varying $P_{0}$ in the range $-mc\leq P_{0}\leq mc$
and superposing the momentum distribution function for every $P_{0}$.
Hereafter, we use $P_{0}$ as the representative momentum for the
discrete subset of momenta connected to the $P_{0}$, and labeled
the subset with $P_{0}$.

Since the following condition is satisfied in the hydrodynamic regime,
(see Eqs. (\ref{eq:range of homogeniety}) and (\ref{eq:relaxation time}))
\begin{equation}
  |k|\sim
   L_{{\rm h}}^{-1}\ll L_{{\rm rel}}^{-1}=
    \lambda_{{\rm gap}}/\langle v\rangle,
  \label{eq:hydrodynamic regime}
\end{equation}
the first term of Eq. (\ref{eq:hydrodynamic approximation of collision op}) (i.e. the flow term)
is much smaller than the collision term ${\cal K}_P^{(0)}$.
Therefore we can treat the flow
term as a perturbation to the collision term.

As shown in Eq. (\ref{eq:discrete momentum states}),
the collision operator acts separately
in each irreducible subspace represented by a momentum $P_{0}$.
In addition, the flow term has only diagonal matrix elements with respect to the momentum $P$.
Therefore we can treat the eigenvalue problem of the collision operator in each $(k,P_0)$-subspace
consisting of a set of states $|k,P_{\nu}(P_{0})\rangle\!\rangle$ ($\nu$ is any integer).
Hence, the eigenstate $\phi_{j}^{(k)}(P)$ can be specified by $P_0$ and $(P_{\nu})$
as {[}cf. Eqs. (\ref{eq:k=00003D0right eigenfunction}) and (\ref{eq:k=00003D0left eigenfunction}){]}
\begin{align}
  \langle\!\langle k,P|\phi_{P_{0};j}^{(k)}\rangle\!\rangle & \equiv
   \sum_{\nu}\phi_{P_{0};j}^{(k)}(P_{\nu})\delta(P-P_{\nu}(P_{0})),
   \nonumber\\
   \langle\!\langle\tilde{\phi}_{P_{0};j}^{(k)}|k,P\rangle\!\rangle & \equiv
   \sum_{\nu}\tilde{\phi}_{P_{0};j}^{(k)}(P_{\nu})\delta(P-P_{\nu}(P_{0})).
\end{align}

The right and left eigenvectors are related by the following relation
\begin{equation}
\tilde{\phi}_{P_{0};j}^{(k)}(P_{\nu})\equiv[\varphi_{P_{0}}^{{\rm eq}}(P_{\nu})]^{-1}\phi_{P_{0};j}^{(k)}(P_{\nu}),
\end{equation}
where $\varphi_{P_{0}}^{{\rm eq}}(P)$ is an equilibrium distribution
given by
\begin{equation}
    \varphi_{P_{0}}^{{\rm eq}}(P_{\nu})\equiv
     \frac{\exp[-\varepsilon_{P_{\nu}}/k_{\mathrm{B}}T]}
     {\sum_{\mu=-\infty}^{\infty}\exp[-\varepsilon_{P_{\mu}}/k_{\mathrm{B}}T]}.
    \label{eq:equilibrium distribution}
\end{equation}

In the hydrodynamic regime (\ref{eq:hydrodynamic regime}),
the eigenvalues of the collision operator in a $(k,P_0)$-subspace can be obtained in perturbation theory as
\begin{equation}
  z_{P_{0};j}^{(k)}=
  z_{P_{0};j}+kz_{P_{0};j}^{[1]}+k^{2}z_{P_{0};j}^{[2]}+\cdots,
  \label{eq:eigenvalue expansion}
\end{equation}
and the eigenvectors are
\begin{align}
  \phi_{P_{0};j}^{(k)}(P_{\nu}) & =
  \phi_{P_{0};j}(P_{\nu})+k\phi_{P_{0};j}^{[1]}(P_{\nu})+k^{2}\phi_{P_{0};j}^{[2]}(P_{\nu})+\cdots,
  \nonumber\\
  \tilde{\phi}_{P_{0};j}^{(k)}(P_{\nu}) & =
  \tilde{\phi}_{P_{0};j}(P_{\nu})+k\tilde{\phi}_{P_{0};j}^{[1]}(P_{\nu})+k^{2}\tilde{\phi}_{P_{0};j}^{[2]}(P_{\nu})+\cdots,
  \label{eq:left eigenvector expansion}
\end{align}
where
the first terms $z_{P_{0};j}$, $\phi_{P_{0};j}(P_{\nu})$ and $\tilde{\phi}_{P_{0};j}(P_{\nu})$ in these equation
are the eigenvalue and eigenvectors of the collision operator in a $(k=0,P_0)$-subspace given at Appendix A.

The coefficient of the $k$-linear term in Eq. (\ref{eq:eigenvalue expansion})
gives the hydrodynamic sound velocity of the exciton $\sigma(P_0)$,
and the coefficient of the $k^2$-term in Eq. (\ref{eq:eigenvalue expansion})
gives the spatial decay rate, i.e., the diffusion coefficient $D(P_0)$:
\begin{equation}
  \sigma(P_0) \equiv z_{P_0;0}^{[1]},\ \ \ \
  D(P_0) \equiv iz_{P_0;0}^{[2]}.
\end{equation}

Using the perturbation theory, these transport coefficients are given by
\begin{align}
  \sigma(P_0) & =
  \langle\!\langle\tilde{\phi}_{P_{0};0}|\frac{\hat{P}}{m}|\phi_{P_{0};0}\rangle\!\rangle\nonumber\\
 & =
 \sum_{\nu=-\infty}^{\infty}
 \frac{P_{\nu}(P_{0})}{m}\phi_{P_{0};0}(P_{\nu}),
  \label{eq:sound velocity}
\end{align}
and
\begin{align}
  D(P_0) & =
  i\sum_{j\neq0}\langle\!\langle\tilde{\phi}_{P_{0};0}|\frac{\hat{P}}{m}|\phi_{P_{0};j}\rangle\!\rangle\frac{1}{-z_{P_{0};j}}\langle\!\langle\tilde{\phi}_{P_{0};j}|\frac{\hat{P}}{m}|\phi_{P_{0};0}\rangle\!\rangle\nonumber\\
 & =
 i\sum_{j\neq0}\frac{1}{-z_{P_{0};j}}\left[\sum_{\nu=-\infty}^{\infty}\frac{P_{\nu}(P_{0})}{m}\phi_{P_{0};j}(P_{\nu})\right]^{2},
  \label{eq:diffusion coefficient}
\end{align}
where the momentum operator $\hat{P}$ is defined in the space
spanned by the basis vectors and satisfies
\begin{equation}
  \langle\!\langle k,P|
   \hat{P}
  |k,P^{\prime}\rangle\!\rangle=
  P\delta(P-P^{\prime}).
\end{equation}

The sound velocity and the diffusion coefficient are defined for each irreducible subset of momenta,
and the momenta in a subset are considered to share the values of the hydrodynamic  sound velocity and the diffusion coefficient as
\begin{equation}
  \sigma(P_\nu)  =\sigma(P_0), \ \ \
  D(P_\nu)  =D(P_0),
  \label{eq:periodicity of diffusion coefficient}
\end{equation}
for all the integers $\nu$.
Since the representative $P_{0}$ of the subsets of momenta takes continuous values in the range $-mc\le P_0\le mc$,
the whole of the momenta $P_{\nu}$ run continuously from $-\infty$ to $\infty$.
Therefore, Eqs.
(\ref{eq:periodicity of diffusion coefficient})
are regarded as defining continuous and periodic functions of momentum $P$.

According to Eqs. (\ref{eq:discrete momentum states}), (\ref{eq:equilibrium distribution}),
 (\ref{eq:sound velocity}), and
 \begin{equation}
\phi_{P_{0};0}(P_{\nu})=\varphi_{P_{0}}^{{\rm eq}}(P_{\nu}),\label{eq:zero right eigenvector}
\end{equation}
which is the equilibrium state belonging to the zero eigenstate of ${\cal K}_P^{(0)}$,
the sound velocity has an explicit expression:
\begin{equation}
 \sigma(P_0,T)=\frac{\displaystyle\sum_{\nu=-\infty}^\infty\frac{P_\nu}{m}\exp\left(-\frac{P_\nu^2}{2mk_\mathrm{B}T}\right)}{\displaystyle\sum_{\mu=-\infty}^\infty\exp\left(-\frac{P_\mu^2}{2mk_\mathrm{B}T}\right)}. \label{eq:sigma_sum}
\end{equation}
It should be emphasized that the hydrodynamic sound mode is obtained though only the zero eigenstate of ${\cal K}_P^{(0)}$, which is in contrast the Diffusion coefficient (\ref{eq:diffusion coefficient}).

The summations in Eq. \eqref{eq:sigma_sum} can be implemented to obtain an analytical expression:
\begin{widetext}
\begin{equation}
\sigma(P_{0},T)
= -\frac{k_{{\rm B}}T}{8mc}
\frac{
      \frac{\partial}{\partial\alpha}
      \vartheta_{3}\bigl(
             \alpha\pi, \tilde{q}_{\scalebox{0.5}{$T$}}
               \bigr)
\big|_{\alpha=\frac{{\tilde P}_0 }{4}}
  -\frac{\partial}{\partial\alpha}
       \vartheta_{3}\bigl(
          \alpha\pi,  \tilde{q}_{\scalebox{0.5}{$T$}}
                \bigr)
            \big|_{\alpha=\frac{1}{2} (\frac{{\tilde P}_0 }{2}-1)}
}
{\vartheta_{3}\left(\frac{\pi{\tilde P}_0 }{2},
      \tilde{q}^4_{\scalebox{0.5}{$T$}}\right)
},
\label{eq: the sound velocity expressed by the theta function}
\end{equation}
\end{widetext}
where
\begin{equation}
  {\tilde P}_0 \equiv \frac{P_0}{mc}, \ \ \ \ \
   \tilde{q}_{\scalebox{0.5}{$T$}}\equiv \exp \left[  -\frac{ \pi^2 }{8}
   \frac{k_{\rm B} T}{ mc^2} \right],
\end{equation}
and $\vartheta_{3}(\tilde{z},\tilde{q})$ is an elliptic theta function
defined as
\begin{equation}
  \vartheta_{3}(\tilde{z},\tilde{q})\equiv
  1+2\sum_{n=1}^{\infty}\tilde{q}^{n^2}\cos(2n\tilde{z}).
  \label{eq:definition of the elliptic theta function}
\end{equation}
Since the $n$-th term of the infinite series in the elliptic theta function
is proportional to the $n^{2}$-th power of $\tilde{q}$,
the elliptic theta function converges rapidly when $\tilde{q}<1$.
Then the one in the denominator of Eq. (\ref{eq: the sound velocity expressed by the theta function}) is expressed as $\tilde{q}_{\scalebox{0.5}{$T$}}^{4}$.

Both of the arguments are less than $1$ at arbitrary temperatures except for $T=0$.
Furthermore, the higher the temperature $T$,
the smaller the arguments and the more rapidly the theta functions converge.

In the temperature range $1\lesssim k_{{\rm B}}T/mc^{2}\lesssim2$
of our interest mentioned in Sec. II, the arguments satisfy the following
relation
\begin{equation}
0<\tilde{q}_{\scalebox{0.5}{$T$}}^{4}\ll1,
\end{equation}
 Therefore, the theta functions in the numerator of Eq. (\ref{eq: the sound velocity expressed by the theta function})
can be approximated by retaining the first two terms as
\begin{equation}
 \vartheta_{3}\left(\alpha\pi,\tilde{q}_{\scalebox{0.5}{$T$}}\right)=
 1+2\tilde{q}_{\scalebox{0.5}{$T$}}\cos(2\alpha\pi)+
 O\left(\tilde{q}_{\scalebox{0.5}{$T$}}^{4}\right),
\label{eq:approx theta fnc 1}
\end{equation}
while the theta function in the denominator of Eq. (\ref{eq: the sound velocity expressed by the theta function}) can be approximated by retaining only the first term as
\begin{equation}
\vartheta_{3}\left(\frac{{\pi\tilde P}_0}{2},\tilde{q}_{\scalebox{0.5}{$T$}}^{4}\right)=
1+O\left(\tilde{q}_{\scalebox{0.5}{$T$}}^{4}\right).
\label{eq:approx theta fnc 2}
\end{equation}
 Substituting Eqs. (\ref{eq:approx theta fnc 1}) and (\ref{eq:approx theta fnc 2})
into Eq. (\ref{eq: the sound velocity expressed by the theta function}),
we obtain an approximate expression for the sound velocity:
\begin{equation}
  \sigma(P_{0},T)=
   \frac{\pi k_{{\rm B}}T}{mc}
    \left\{
       e^{ -\frac{ \pi^2 k_{\rm B} T } {8mc^2} }
        \sin\left(\frac{\pi P_0 }{2mc}\right)+
       O\left(\tilde{q}_{\scalebox{0.5}{$T$}}^{5}\right)
    \right\}.
  \label{eq:approximated sound velocity}
\end{equation}

Equation (\ref{eq:approximated sound velocity}) shows that the sound
velocity is the largest when $P_{0}/mc=1$ and it goes to zero at
high temperatures.
\setlength\textfloatsep{-10pt}

\begin{center}

\begin{figure}[htbp]
\includegraphics[scale=0.6]
{
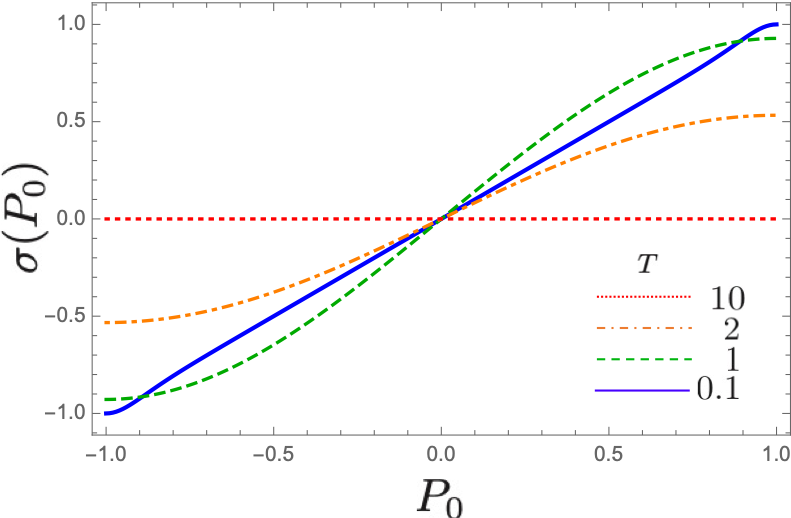}
\caption
 {
 \label{fig:Momentum dependence of the sound velocity}
 Momentum dependence of the sound velocity of a hydrodynamic mode of the exciton at temperatures
 $T=0.1$(solid), $1$(dashed),$2$(dash-dotted),
 $10$(dotted) in units where $m=1,c=1,\hbar=1,k_{{\rm B}}=1$.
 }
\end{figure}
\begin{figure}[htbp]
\includegraphics[scale=0.6]
{
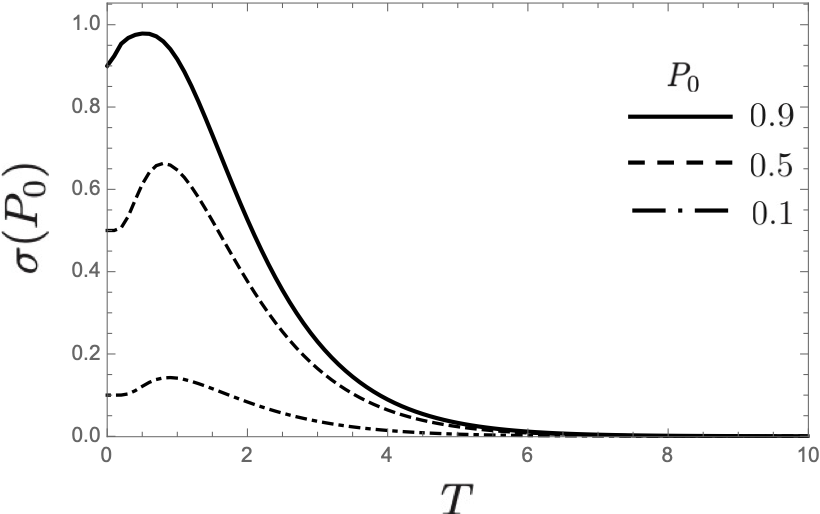}
\caption
 {
 \label{fig:Temperature dependence of the sound velocity}
 Temperature
 dependence of the sound velocity of a hydrodynamic mode of the exciton
 for momenta $P_{0}=0.9$ (solid), $0.5$ (dashed), $0.1$ (dash-dotted) in units where $m=1,c=1,\hbar=1,k_{{\rm B}}=1$.
 }
\end{figure}
\vspace*{7mm}
\end{center}

In Figs. \ref{fig:Momentum dependence of the sound velocity} and
\ref{fig:Temperature dependence of the sound velocity}, we show the
momentum dependence and the temperature dependence of the sound velocity
(\ref{eq: the sound velocity expressed by the theta function}).

In
these figures, we show the results for our interesting  temperature  domain
$1\lesssim k_{{\rm B}}T/mc^{2}\lesssim2$, in addition to the temperature  domain
 at lower and higher temperatures.

Note that the temperatures $k_{{\rm B}}T/mc^{2}=0.1$, $1$, $2$ and $10$
correspond to $T\simeq22{\rm K}$, $220{\rm K}$, $440{\rm K}$ and
$2200{\rm K}$, respectively. It can be confirmed that the sound velocity
is the largest at $P_{0}/mc=1$ and it goes to zero at high temperatures
as shown in Eq. (\ref{eq:approximated sound velocity}). Furthermore,
Fig. \ref{fig:Temperature dependence of the sound velocity} shows
that the sound velocity has relatively large values in the temperature
 domain $1\lesssim k_{{\rm B}}T/mc^{2}\lesssim2$.
As a result,  one can readily see the momentum dependence of the sound velocity
in this temperature domain.

\begin{center}
\begin{figure}[htbp]
\includegraphics[scale=0.6]
{
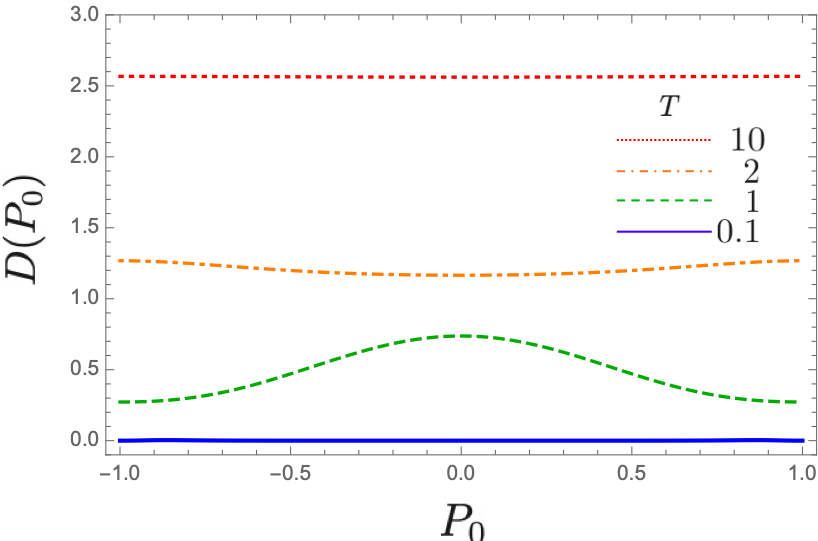}
\setlength\abovecaptionskip{1.1pt}
\caption
 {
 \label{fig:Momentum dependence of the diffusion coefficient}
 Momentum dependence of the diffusion coefficient of the exciton at temperatures
 $T=0.1$(solid), $1$(dashed),$2$(dash-dotted),
 $10$(dotted) in units where $m=1,c=1,\hbar=1,k_{{\rm B}}=1\ {\rm and}\ |\lambda_{\infty}|=1$.
 }
\end{figure}
\vspace*{-1mm}
\begin{figure}[htbp]
\includegraphics[scale=0.6]
{
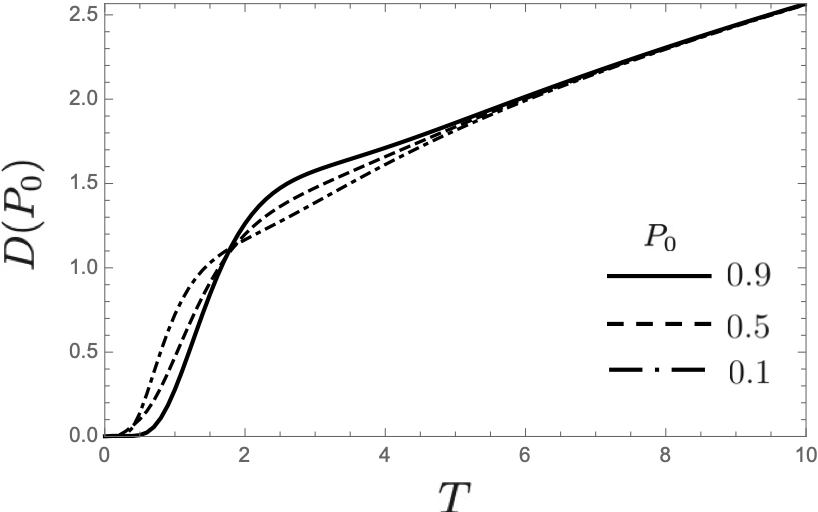}
\caption
 {
 \label{fig:Temperature dependence of the diffusion coefficient}
 Temperature dependence of the diffusion coefficient of the exciton for momenta
 $P_{0}=0.9$ (solid), $0.5$ (dashed), $0.1$ (dash-dotted) in units where $m=1,c=1,\hbar=1,k_{{\rm B}}=1\ {\rm and}\ |\lambda_{\infty}|=1$.
 }
\end{figure}

\end{center}
\setlength\textfloatsep{10pt}

Let us now consider the diffusion coefficient (\ref{eq:diffusion coefficient})
expressed in terms of the nonzero eigenvalues $z_{P_{0};j}$
and corresponding right eigenvectors $\phi_{P_{0};j}(P_{\nu})$ of the collision operator ${\cal K}_P^{(0)}$.
In this paper, we present the results obtained by the numerical calculation for these eigenvalues and the eigenvectors.
The analytic form of the solution of the eigenvalue problem will be presented elsewhere.
Here, we show the numerical results for the momentum dependence
and the temperature dependence of the diffusion coefficient (\ref{eq:diffusion coefficient})
in Figs. \ref{fig:Momentum dependence of the diffusion coefficient}
and \ref{fig:Temperature dependence of the diffusion coefficient}.

It is found that the momentum dependence of the diffusion coefficient is significant in the temperature domain of our
our interest $1\lesssim k_{{\rm B}}T/mc^{2}\lesssim2$ (see Fig. \ref{fig:Momentum dependence of the diffusion coefficient}).
For extremely high and low temperatures, the momentum dependence is no significant.

\subsection{B. Convection-diffusion equation }

In this section, we consider the time evolution of the Wigner distribution function for the exciton, and drive the convection-diffusion equation.

Since the collision operator is defined in each disjoint subspace associated with a subset of momenta,
the time evolution of the distribution function is also determined for each disjoint subspace.
The time evolution of the Fourier component of the Wigner distribution function $f_{k}(P_{\nu},t)$
that belongs to a certain momentum subset represented by $P_{0}$
is given by the eigenfunction expansion method as
\begin{equation}
  f_{k}(P_{\nu},t)=
   \sum_{j}
   e^{-iz_{P_{0};j}^{(k)}t}\phi_{P_{0};j}^{(k)}(P_{\nu})
    \langle\!\langle\tilde{\phi}_{P_{0};j}^{(k)}|f(t=0)\rangle\!\rangle,
  \label{eq:fk distribution functioin}
\end{equation}
 where
\begin{equation}
  \langle\!\langle
   \tilde{\phi}_{P_{0};j}^{(k)}|f(t=0)
  \rangle\!\rangle=
   \sum_{\mu=-\infty}^{\infty}
   \tilde{\phi}_{P_{0};j}^{(k)}(P_{\mu})f_{k}(P_{\mu},0).
  \label{eq:left eigenvectors and initial state}
\end{equation}
Eqs. (\ref{eq:fk distribution functioin}) and (\ref{eq:left eigenvectors and initial state})
are the functions defined for each value of $P_0$.
However, since $P_{\nu}$ and $P_{\mu}$ take any real number when $P_{0}$ varies continuously in the domain $-mc\le P_0\le mc$,
the function $f_{k}(P,t)$ is defined to be a continuous function of $P$.

After the momentum relaxation $t\gtrsim\tau_{{\rm rel}}$,
the system has reached the local equilibrium,
\begin{align}
  f_{k}(P,t) & \xrightarrow{t\gtrsim\tau_{{\rm rel}}}
   e^{-iz_{P;0}^{(k)}t}\phi_{P;0}^{(k)}(P)
    \langle\!\langle\tilde{\phi}_{P;0}^{(k)}|f(t=0)\rangle\!\rangle.
  \label{eq:distribution fnc after local eq}
\end{align}

\begin{widetext}
In the hydrodynamic regime (\ref{eq:hydrodynamic regime}), the function $f_{k}(P,t)$ is expressed as
\begin{equation}
  f_{k}(P,t\gtrsim\tau_{{\rm rel}})\simeq
   e^{-ik\sigma(P)t}e^{-k^{2}D(P)t}\phi_{P;0}^{(k)}(P)
    \langle\!\langle\tilde{\phi}_{P;0}^{(k)}|f(t=0)\rangle\!\rangle,
  \label{eq:distribution fnc after local eq in the hydrodynamic regime}
\end{equation}
 where $z_{P;0}^{(k)}$ is approximated up to the second-order of
$k$ in Eq. (\ref{eq:eigenvalue expansion}).
The Wigner distribution function for the exciton defined by Eq. (\ref{eq:fourier transform to Wigner distribution function})
in the hydrodynamic regime is then given by
\begin{equation}
  f^{W}(X,P,t\gtrsim\tau_{{\rm rel}})\simeq
   \frac{1}{2\pi}\int_{-\infty}^{\infty}\!\!\!\!dk\
    e^{ik(X-\sigma(P)t)}e^{-k^{2}D(P)t}\phi_{P;0}^{(k)}(P)
     \langle\!\langle\tilde{\phi}_{P;0}^{(k)}|f(t=0)\rangle\!\rangle.
  \label{eq:formal solution of the Wigner distribution function after local equiribrium}
\end{equation}
Differentiating this expression
with respect to $X$ and $t$, we then obtain a convection-diffusion
equation for $t\gtrsim\tau_{{\rm rel}}$ with a sound velocity $\sigma(P)$
and a diffusion coefficient $D(P)$ as

\begin{equation}
  \frac{\partial}{\partial t}f^{W}(X,P,t)=
   -\sigma(P)\frac{\partial}{\partial X}f^{W}(X,P,t)+
   D(P)\frac{\partial^{2}}{\partial X^{2}}f^{W}(X,P,t).
  \label{eq:convection-diffusion equation}
\end{equation}
\end{widetext}

We note that the effects of  the sound velocity and the momentum dependence of the diffusion coefficient
are significant. However, this convection-diffusion equation reduces to the usual diffusion equation for
a high temperature, since the sound velocity is almost zero and the diffusion coefficient reaches a constant that does not depend on momentum at a high temperature.

\hypertarget{IV}{}
\section{phenomenological diffusion coefficient}

In this section, we consider the phenomenological diffusion coefficient $D^{(x)}(t)$ in Eq. (\ref{eq:phenomenological diffusion coefficient}).
In that expression, the symbol $\langle\cdots\rangle_{t}$ indicates to take the average
over the Wigner distribution function $f^{W}(X,P,t)$ as
\begin{align}
\langle g(X,P)\rangle_{t} & \equiv\int_{-\infty}^{\infty}\!\!\!\!\!\!\!dX\int_{-\infty}^{\infty}\!\!\!\!\!\!\!dP\ g(X,P)\ f^{W}\!(X,P,t)\nonumber\\
 & =\int_{-\infty}^{\infty}\!\!\!\!\!\!\!dX\int_{-mc}^{mc}\!\!\!\!\!\!\!\!dP_{0}\sum_{\nu=-\infty}^{\infty}\ g(X,P_{\nu})\ f^{W}\!(X,P_{\nu},t),
 \label{eq:difinition of average 2}
\end{align}
where $g(X,P)$ is an arbitrary function of $X$ and $P$.
The integration over $P$ can be replaced with the integration of
$P_{0}$ from $-mc$ to $mc$ after the summation over all the discrete momenta $P_{\nu}$ connected to each $P_{0}$.

As explained in the introduction, there is a numerical result reported by Pouthier \cite{2009VPouthier}
that the phenomenological diffusion coefficient (\ref{eq:phenomenological diffusion coefficient}) for
the exciton increases linearly with time and diverges in the long-time limit.
We
now
show that the linear time dependence of $D^{(x)}(t)$ can be understood using an analytic expression for $D^{(x)}(t)$
in terms of the transport coefficients in the kinetic equation,
and that the divergence of $D^{(x)}(t)$ in the long-time limit is due to the phase mixing
which arises from the momentum dependence of the sound velocity.

In order to show this, we first consider a special case where
the initial condition is given by a
pure state associated to a
wave function with
the minimum uncertainty
\begin{equation}
  \psi_0(X)=
  \left[\frac{1}{2\pi(\Delta X)^{2}}\right]^{\frac{1}{4}}
   \exp\left[i\frac{P^{\prime}}{\hbar}X-\frac{X^{2}}{4(\Delta X)^{2}}\right],
  \label{eq: initital wave function given as minimum uncertainty}
\end{equation}
where $P^{\prime}$ is a peak position of the initial momentum distribution.
We note that Pouthier gave an initial condition as a delta function
of $X$. Therefore, his initial condition corresponds to the case
$\Delta X\rightarrow0$ in our initial condition.

For this state (\ref{eq: initital wave function given as minimum uncertainty}),
the Fourier component of the Wigner distribution function at $t=0$ is given by (see Eq. (\ref{eq:Wigner expression}))
\begin{align}
  f_{k}(P,t=0) & =
   \left( P+\frac{\hbar k}{2}\right|f(0)\left|P-\frac{\hbar k}{2}\right)
   \nonumber\\
  & =
   \left( P+\frac{\hbar k}{2}\middle|\psi_0\right)
   \left( \psi_0\middle|P-\frac{\hbar k}{2}\right)\nonumber\\
  & =
   \frac{1}{\sqrt{2\pi(\Delta P)^{2}}}
    \exp\left[-\frac{(P-P^{\prime})^{2}}{2(\Delta P)^{2}}-\frac{(\Delta X)^{2}}{2}k^{2}\right],
  \label{eq:initial fourier component of Wigner distribution function}
\end{align}
 where
\begin{equation}
  \Delta X\cdot\Delta P=\hbar/2.
\end{equation}
Substituting Eq. (\ref{eq:initial fourier component of Wigner distribution function})
into Eqs. (\ref{eq:left eigenvectors and initial state}) and (\ref{eq:distribution fnc after local eq in the hydrodynamic regime}),
we obtain the Fourier component of the Wigner distribution function for $t\gtrsim\tau_{{\rm rel}}$ as
\begin{widetext}
\begin{equation}
  f_{k}(P_{\nu},t\gtrsim\tau_{{\rm rel}})\simeq
   e^{-\{ik\sigma(P_{0})+k^{2}D(P_{0})\}t}\phi_{P_{0};0}^{(k)}(P)
    \sum_{\mu=-\infty}^{\infty}
     \tilde{\phi}_{P_{0};0}^{(k)}(P_{\mu})
     \frac{1}{\sqrt{2\pi(\Delta P)^{2}}}
     \exp\left[-\frac{(P_{\mu}-P^{\prime})^{2}}{2(\Delta P)^{2}}-\frac{(\Delta X)^{2}}{2}k^{2}\right].
  \label{eq:the Fourier component of the Wigner distribution function in local eq}
\end{equation}
Then, we approximate  $\phi_{P_{0};0}^{(k)}(P)$ and $\tilde{\phi}_{P_{0};0}^{(k)}(P)$ by
 the eigenfunctions in the $k=0$ subspace $\phi_{P_{0};0}(P)$
and $\tilde{\phi}_{P_{0};0}(P)$, respectively, for the hydrodynamic case with a small value of
$k$ (see Eqs. (\ref{eq:left eigenvector expansion})), and obtain
%
\begin{equation}
  f_{k}(P_{\nu},t\gtrsim\tau_{{\rm rel}})\simeq
   e^{-\{ik\sigma(P_{0})+k^{2}D(P_{0})\}t}\varphi_{P_{0}}^{{\rm eq}}(P_{\nu})
    \sum_{\mu=-\infty}^{\infty}
     \frac{1}{\sqrt{2\pi(\Delta P)^{2}}}
     \exp\left[-\frac{(P_{\mu}-P^{\prime})^{2}}{2(\Delta P)^{2}}-\frac{(\Delta X)^{2}}{2}k^{2}\right],
  \label{eq:explicit Fourier component at local eq}
\end{equation}
where we have used the relations (\ref{eq:zero right eigenvector}) and (\ref{eq:zero left eigenvector}).
We note that is a mixed state in spite of the fact that the initial condition was a pure state.

By performing a Fourier transform on Eq. (\ref{eq:explicit Fourier component at local eq}),
the time evolution of the Wigner distribution function for $t\gtrsim\tau_{{\rm rel}}$ is obtained as
\begin{equation}
  \begin{split}
    f^{W}(X,P_{\nu},t\gtrsim\tau_{{\rm rel}}) & =
      \frac
       {\exp\left(-\frac{P_{\nu}^{2}}{2mk_{{\rm B}}T}\right)}
       {\sum_{\kappa=-\infty}^{\infty}
        \exp\left(-\frac{P_{\kappa}^{2}}{2mk_{{\rm B}}T}\right)}
      \sum_{\mu=-\infty}^{\infty}
       \frac{1}{\sqrt{2\pi(\Delta P)^{2}}}
       \exp\left[-\frac{(P_{\mu}-P^{\prime})^{2}}{2(\Delta P)^{2}}\right]\\
    & \times
      \frac{1}{\sqrt{2\pi\{(\Delta X)^{2}+2D(P_{0})t\}}}
      \exp\left[-\frac{(X-\sigma(P_{0})t)^{2}}{2\{(\Delta X)^{2}+2D(P_{0})t\}}\right],
  \end{split}
\label{eq:explicit form of the Wigner distribution function of the exciton}
\end{equation}
\end{widetext}
where $P_{\nu}$, $P_{\mu}$ and $P_\kappa$ are discrete momenta
belonging to the subset of momenta represented by $P_{0}$.

Now we can calculate the average $\langle\cdots\rangle_{t}$ in Eq. (\ref{eq:phenomenological diffusion coefficient}).
Since the initial spatial distribution is given as a Gaussian,
one can integrate over $X$ by using the formula for the Gaussian integral.
After the integration over $X$, the phenomenological
diffusion coefficient (\ref{eq:phenomenological diffusion coefficient})
can be expressed as Eq. (\ref{eq:phenomenological diffusion coefficient in 1D system}) (see Appendix B for a derivation).
The explicit form of the averages
$\langle\cdots\rangle_{\rm eq}$ in Eq. \eqref{eq:phenomenological diffusion coefficient in 1D system} are given by \eqref{eq:<D(P)>_ell} and \eqref{eq:<sigma(P)>_ell}.

Furthermore, in this specific system, the averages $\langle\cdots\rangle_{\rm eq}$ of the transport coefficients
are equal to the averages of them over
the initial Wigner distribution function, such as,
\begin{align}
  \bigl\langle D(P)\bigr\rangle_{\rm eq}&=
  \bigl\langle D(P)\bigr\rangle_{t=0},\nonumber\\
  \bigl\langle \sigma(P)\bigr\rangle_{\rm eq}&=
  \bigl\langle \sigma(P)\bigr\rangle_{t=0}.
  \label{eq:<>_ell=<>_0}
\end{align}

The reason that Eqs. \eqref{eq:<>_ell=<>_0} satisfy is as follows:
Relaxation of momentum distribution occurs only among the momenta in each subset,
and hence the sum of the momentum distribution probability within each subspace is conserved during the momentum equilibration.
Besides, values of the transport coefficients $\sigma(P_0)$ and $D(P_0)$
are shared by the momenta in the subset connected to the momentum $P_0$ via the collision operator.
Therefore, the average of the transport coefficients in each momentum subspace is conserved during the momentum equilibration.

For more general initial condition, one can obtain $D^{(x)}(t)$ as Eq. (\ref{eq:phenomenological diffusion coefficient in 1D system}) with a time-independent extra term which comes from the deviation of the initial distribution from Gaussian.
The proof is given in Appendix C, where we use
the theorem
that any square integrable function can be approximated with arbitrary precision by
the linear combination of Gaussian \cite{2008CCalcaterra,2008CCalcaterraABoldt}.

The expression Eq. (\ref{eq:phenomenological diffusion coefficient in 1D system}) shows that the
phenomenological diffusion coefficient (\ref{eq:phenomenological diffusion coefficient}) consists of the two parts:
the first term is due to the diffusion process and the second term is due to the phase mixing.
The diffusion process is an irreversible process associated with entropy production.
On the other hand,
the phase mixing is a reversible process in which the wave packet spreads along the spatial direction
in the phase space because of the difference of the sound velocity according to momentum.
Therefore, the spreading of the spacial distribution occurs owing to completely different two mechanisms.
The divergence of the phenomenological diffusion coefficient (\ref{eq:phenomenological diffusion coefficient})
is ascribed to the phase mixing.

In Fig. \ref{fig:Time evolution of the Wigner distribution function} is shown
the time evolution of the Wigner distribution function (\ref{eq:explicit form of the Wigner distribution function of the exciton})
for $t > \tau_{\rm rel}$.
The distribution in Fig. \ref{fig:Time evolution of the Wigner distribution function}(a)
has side peaks at $P_{1}/mc=1.5$ and $P_{-1}/mc=-2.5$ besides the main peak at $P_{0}/mc=P^{\prime}/mc=0.5$.
This is because the exciton can make transitions only within the momentum subset
(\ref{eq:discrete momentum states}) connected to the initial momenta.

After the equilibrium state for the momentum is established,
the components of the Wigner distribution function
at momenta (\ref{eq:discrete momentum states})
connecting
to a $P_{0}$
move with the same velocity $\sigma(P_{0})$.
Moreover, for those momenta
the variance along the $X$-axis increases
at the same rate of the diffusion coefficient $D(P_{0})$.

In Fig. \ref{fig:Cross sections of the Wigner distribution function}, we show
the components of the Wigner distribution function at the momenta
connecting
to a $P_{0}/mc=0.5$ and their projection onto the plane perpendicular to the $P$-axis.
Those figures illustrate the fact that these momenta belonging to the same momentum subset share the same hydrodynamic sound velocity and the same diffusion coefficient.

We show in Fig. \ref{fig:Cross sections of the Wigner distribution function at different momentum subspace}
the cross sections of the components of the Wigner distribution function at momenta
belonging to different momentum subset. One can see that the diffusion
processes broaden the variances of each cross sections with rates
of the different diffusion coefficients. In addition, the difference
between the peak positions of each cross sections increases with time
since these cross sections of components of the Wigner distribution function move with different sound velocities
associated with the different value of the momentum.

Hence if we observe the spatial distribution of the exciton defined by
\begin{equation}
  f(X,t)\equiv
   \int_{-\infty}^{\infty}f^{W}(X,P,t)dP,
  \label{eq:spatial distribution of the exciton}
\end{equation}
this function spreads in time not only due to the diffusion processes but also due to the effect of the phase mixing.

\begin{widetext}
\begin{center}
\begin{figure}[htbp]
\centering{}
\includegraphics[scale=0.5]
{
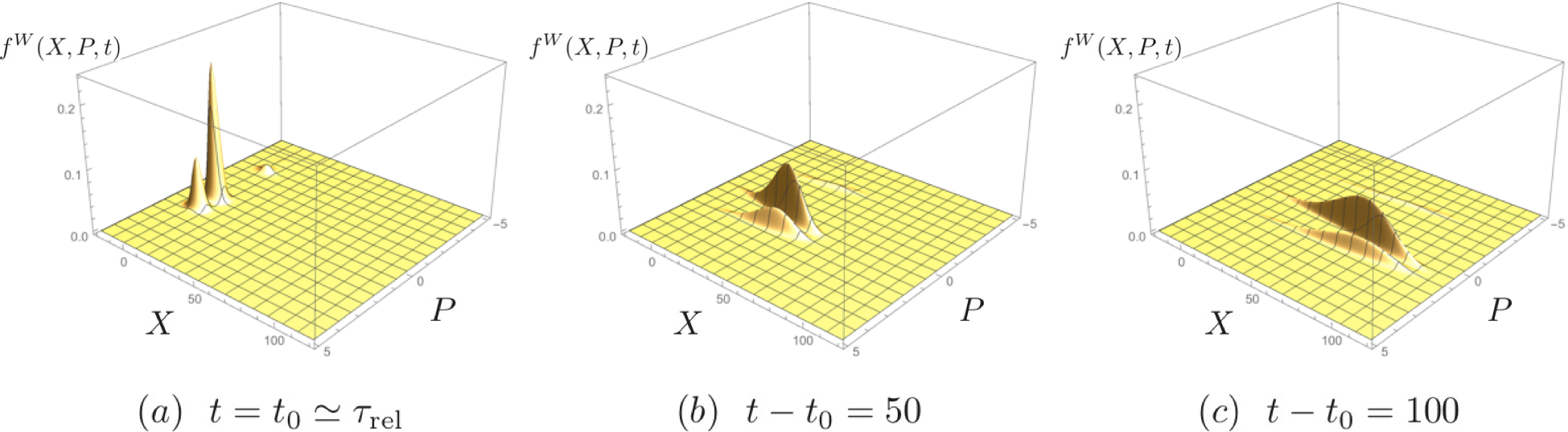}
\caption
 {\label{fig:Time evolution of the Wigner distribution function}
 Time evolution of
 the Wigner distribution function for the exciton for $T=1$, $\Delta X=3$, $P^{\prime}=0.5$
 in units where $m=1,c=1,\hbar=1,k_{{\rm B}}=1\ {\rm and}\ |\lambda_{\infty}|=1$.}
\end{figure}
\end{center}
\begin{figure}[htbp]
\centering{}\includegraphics[scale=0.5]{
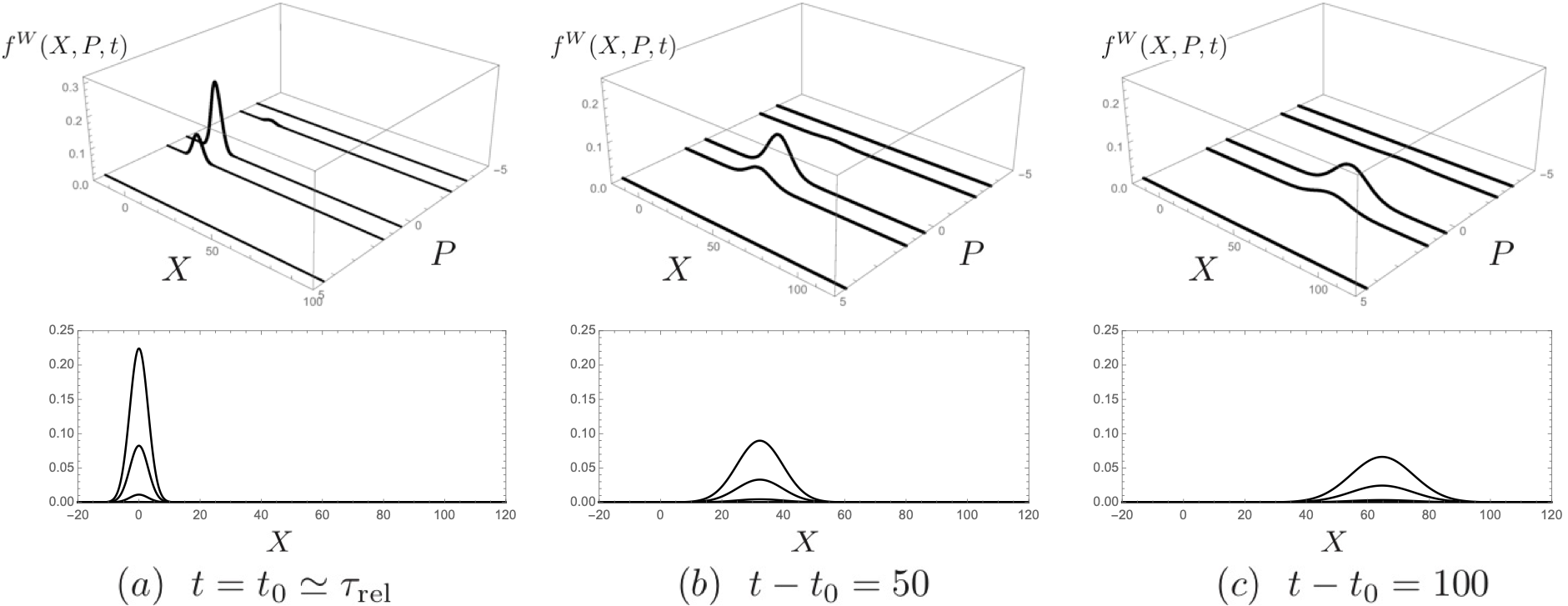}\caption{\label{fig:Cross sections of the Wigner distribution function}Cross sections of
the Wigner distribution function for the exciton shown in Fig.  \ref{fig:Time evolution of the Wigner distribution function}
at $P_{0}=0.5$ in units where $m=1,c=1,\hbar=1,k_{{\rm B}}=1\ {\rm and}\ |\lambda_{\infty}|=1$.
Each section moves with a same velocity $\sigma(P_{0}=0.5)$. The variances along $X$-axis increase at a same rate of a diffusion coefficient
$D(P_{0}=0.5)$.
}
\end{figure}
\begin{center}
\begin{figure}[htbp]
\centering{}\includegraphics[scale=0.5]{
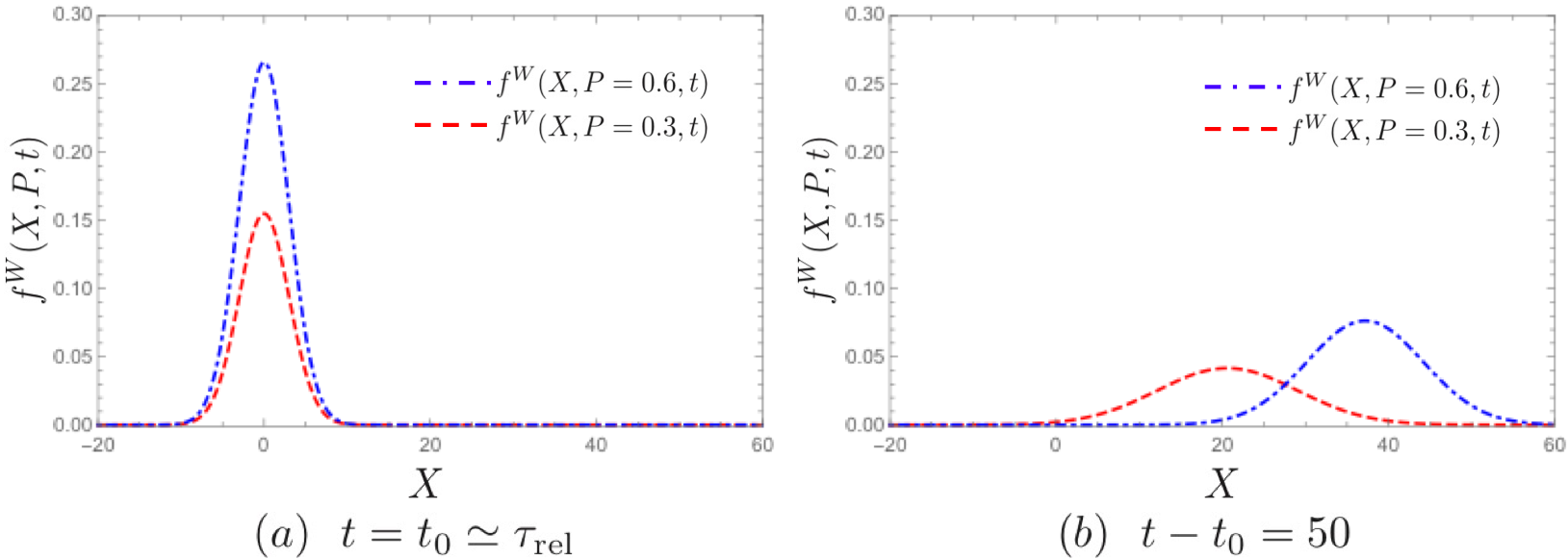}\caption{\label{fig:Cross sections of the Wigner distribution function at different momentum subspace}Cross
sections of the Wigner distribution function for the exciton shown in Fig. \ref{fig:Time evolution of the Wigner distribution function}
at $P=0.3$(dash-dotted), $0.5$(dashed) which belong to different momentum subspaces.
The units $m=1,c=1,\hbar=1,k_{{\rm B}}=1$ and $|\lambda_{\infty}|=1$ are used.}
\end{figure}
\end{center}
\end{widetext}

\hypertarget{V}{}
\section{Conclusions}

We have shown that a hydrodynamic mode emerges
in relaxation processes of an exciton weakly coupled with a thermal phonon field in a 1D molecular chain.
We obtained the hydrodynamic mode in the formalism of the complex spectral analysis of the Liouvillian
by treating the flow term in the effective Liouvillian as a perturbation to the collision term.
The mode is featured by a sound velocity and a diffusion coefficient,
both of which depend on the momentum of the exciton.
As a result, hydrodynamic sound wave propagation and diffusive relaxation coexist,
and the time evolution of the Wigner function of the exciton obeys the convection-diffusion equation.
Phase mixing due to the momentum dependence of the sound velocity leads to anomalous diffusion
in the sense that the increase rate of the mean-square displacement of the exciton increases linearly with time
and diverges in the long-time limit as Eq.~\eqref{eq:phenomenological diffusion coefficient in 1D system}.

One-dimensionality is crucial in giving the system
with the properties mentioned above.
From constraints on the collision processes in 1D represented by the resonance condition,
it follows that the momentum space separates into infinite sets of disjoint subspaces
dynamically independent of one another.
Consequently, momentum relaxation occurs only within each subspace
toward the Maxwell distribution constrained within the subspace.
Thus, the transport coefficients are defined in each irreducible subspace,
and in this sense the sound velocity and the diffusion coefficient are momentum-dependent.
As a result, although the phenomenological diffusion coefficient defined
by Eq.~\eqref{eq:phenomenological diffusion coefficient} diverges in the long-time limit,
the diffusion coefficient as the transport coefficient in the kinetic equation is well-defined.

Moreover, a novel mechanism is responsible for the nonvanishing of the hydrodynamic sound velocity in our 1D system.
As is well known, in classical gas systems, the degeneracy of the collisional invariants
associated with the zero eigenmodes of the collision operator is lifted
by the flow term in the inhomogeneous kinetic equation resulting
in macroscopic hydrodynamic modes such as a sound wave mode and a diffusion mode
in the hydrodynamic regime~\cite{1975RBalescu,1977PResiboisMdeLeenery}.
On the other hand, the appearance of the hydrodynamic mode with nonvanishing sound velocity in our 1D system
is not due to a degeneracy,
but due to a property that equilibration of momentum distribution occurs separately in each subset of momenta.
The momentum distribution function on one of the subsets, in general, is neither even nor odd,
because only one of $P$ and $-P$ is in the subset,
while the velocity of the exciton $P/m$ is an odd function of the momentum.
Thus, the sound velocity, which is given by the average of the velocity of the exciton
over the equilibrium momentum distribution on a subset of momenta (see Eq.~\eqref{eq:sigma_sum}),
is non-vanishing in our 1D system.

When it comes to systems in more than one dimensions, the separation of momenta into subsets does not occur
because of the angular degrees of freedom in collision processes (see above Eq. (\ref{eq:1D discrete momenta})).
As a result, the equilibrium momentum distribution function is an even function.
Hence, the sound velocity vanishes for systems in more than one dimensions.

Some authors have already pointed out that the phenomenological diffusion coefficient
defined by Eq.~\eqref{eq:phenomenological diffusion coefficient} has a linear term with respect to time~\cite{1996HDoldererMWagner,1998HDoldererMWagner,2009VPouthier}.
However, it appears that in those situations the time-dependence of the phenomenological diffusion coefficient
comes from phase mixing in free-particle like motion possibly with renormalization of the mass (polaron effect),
because resonance with phonons is not effective for excitons with narrow excitation energy bandwidth
treated in the papers.
Note that phase mixing in free particle motion may occur also in higher spatial dimensions,
in contrast to the fact that phase mixing due to the momentum-dependent sound velocity of the hydrodynamic mode
appears only in 1D,
as we have discussed in the present paper.

Finally, we emphasize the importance of the effects from the environment at finite temperatures
in understanding the behavior of biological systems.
We hope to clarify the role played by the hydrodynamic mode with non-vanishing sound velocity
in bio-energy transfer processes in the future study.

\begin{acknowledgments}
This work was partially supported by JSPS KAKENHI Grant Number JP17K05585.

\end{acknowledgments}

\appendix

\hypertarget{A}{}
\section{Relaxation modes of the momentum distribution function}

In this appendix, we summarize the time evolution of the momentum distribution
function of the exciton given by
\begin{equation}
 f_{0}(P,t)\equiv\langle\!\langle0,P|f(t)\rangle\!\rangle=(P|f(t)|P),
\end{equation}
presented in our previous papers \cite{2010KKankiSTanakaBATayTPetrosky,2011BATayKKankiSTanakaTPetrosky}
for the weak-coupling case. For details, refer to these papers.

The kinetic equation for the momentum distribution function of the
exciton is written as
\begin{equation}
 i\frac{\partial}{\partial t}{\cal P}^{(0)}|f(t)\rangle\!\rangle=
 \overline{\Psi}_{2}^{(0)}(i0^{+}){\cal P}^{(0)}|f(t)\rangle\!\rangle,
 \label{eq:kinetic eq}
\end{equation}
where the collision operator $\overline{\Psi}_{2}^{(0)}(i0^{+})$ is defined by Eq. (\ref{eq:collison operator}).
The notation ${\cal P}^{(0)}$ in Eq. (\ref{eq:kinetic eq}) denotes
the projection operator to the space spanned by the diagonal elements
of the exciton density matrix with respect to the momentum states,
and defined as
\begin{equation}
 {\cal P}^{(0)}\equiv\int dP|0,P\rangle\!\rangle\langle\!\langle0,P|.
\end{equation}

By multiplying $\langle\!\langle0,P|$ from the left of Eq. (\ref{eq:kinetic eq}),
we obtain
\begin{equation}
 i\frac{\partial}{\partial t}f_{0}(P,t)=
 {\cal K}_P^{(0)}f_{0}(P,t),
 \label{eq:kinetic eq of wigner expression}
\end{equation}
where
${\cal K}_P^{(0)}$ is a difference operator given as Eq. (\ref{eq:concrete form of the collision op}).

Taking into account the consequences of the resonance condition in the collision operator (\ref{eq:concrete form of the collision op}), the
kinetic equation (\ref{eq:kinetic eq of wigner expression}) reduces to a difference
equation, which can be written in a standard form of Markov master
equation with gain and loss terms as
\begin{equation}
  i\frac{\partial}{\partial t}f_{0}(P,t)=\!\!\!
  \sum_{P^{\prime}=-P\pm2mc}\!\!\!
   \left\{
    K_{P,P^{\prime}}f_{0}(P^{\prime},t)-K_{P^{\prime},P}f_{0}(P,t)
   \right\},
  \label{eq:master equation}
\end{equation}
where the sum on the right-hand side is the sum of the two cases,
$P^{\prime}=-P+2mc$ and $P^{\prime}=-P-2mc$, and the transition
probabilities $K_{P,P^{\prime}}/i$ are given by
\begin{equation}
  -iK_{P,P^{\prime}}\equiv
   \frac{g^{2}m\Delta_{0}^{2}}{\hbar^{2}\rho_{M}c}
   \frac{1}
   {|\exp[(\varepsilon_{P}-\varepsilon_{P^{\prime}})/k_{\mathrm{B}}T]-1|}.
\end{equation}

Now, we consider the eigenvalue problem of the collision operator,
\begin{equation}
  {\cal K}_P^{(0)}\phi_{j}(P)=z_{j}\phi_{j}(P).
 \label{eq:k=00003D0eigenvalue problem of the collision operator}
\end{equation}
The collision operator ${\cal K}_P^{(0)}$ is non-Hermitian operator.
Thus, we have to consider the so-called right and left  eigenvalue
problem of the collision operator respectively. Since their eigenvalue
equations consist of components with discrete momenta related to a
$P_{0}$, the right and left  eigenvectors are expressed with the
discrete set of momenta (\ref{eq:discrete momentum states}) as
\begin{align}
  \langle\!\langle0,P|\phi_{P_{0};j}\rangle\!\rangle &\equiv
   \sum_{\nu}\phi_{P_{0};j}(P_{\nu})\delta(P-P_{\nu}(P_{0})),
  \label{eq:k=00003D0right eigenfunction}\\
  \langle\!\langle\tilde{\phi}_{P_{0};j}|0,P\rangle\!\rangle &\equiv
   \sum_{\nu}\tilde{\phi}_{P_{0};j}(P_{\nu})\delta(P-P_{\nu}(P_{0})),
  \label{eq:k=00003D0left eigenfunction}
\end{align}
 where the expression $P_{\nu}(P_{0})$ indicates that $P_{\nu}$
is connected as Eq. (\ref{eq:discrete momentum states}) to a particular
$P_{0}$ .We also introduce a representation of the eigenvectors $|\phi_{j}\rangle\!\rangle_{P_{0}}$
and $_{P_{0}}\langle\!\langle\tilde{\phi}_{j}|$ as vectors with components
on a discrete set of momenta (\ref{eq:discrete momentum states})
\begin{align}
  _{P_{0}}\langle\!\langle0,P_{\nu}|\phi_{j}\rangle\!\rangle_{P_{0}} &
  \equiv
   \phi_{P_{0};j}(P_{\nu}),\\
  _{P_{0}}\langle\!\langle\tilde{\phi}_{j}|0,P_{\nu}\rangle\!\rangle_{P_{0}} & \equiv
   \tilde{\phi}_{P_{0};j}(P_{\nu}),
\end{align}
with the basis vectors $|0,P_{\nu}\rangle\!\rangle_{P_{0}}$ and $_{P_{0}}\langle\!\langle0,P_{\nu}|$
satisfying
\begin{equation}
  _{P_{0}}\langle\!\langle0,P_{\mu}|0,P_{\nu}\rangle\!\rangle_{P_{0}}=
  \delta_{\mu,\nu}^{{\rm Kr}}.
\end{equation}

The right-eigenvalue equation among the components $\phi_{P_{0};j}(P_{\nu})$
can be written as a set of equations,
\begin{widetext}
\begin{equation}
  -(K_{\nu+1,\nu}+K_{\nu-1,\nu})\phi_{P_{0};j}(P_{\nu})+
  K_{\nu,\nu-1}\phi_{P_{0};j}(P_{\nu-1})+
  K_{\nu,\nu+1}\phi_{P_{0};j}(P_{\nu+1})=
   z_{P_{0};j}\phi_{P_{0};j}(P_{\nu}),
  \label{eq:right eigenvalue eq}
\end{equation}
where $\nu=0,\pm1,\pm2,\cdots,$ and $K_{\nu,\mu}\equiv K_{P_{\nu}P_{\mu}}$.
The component $\phi_{P_{0};j}(P_{\nu})$ is the $j$-th right eigenvector
on the set of momenta $P_{\nu}$
with a fixed $P_{0}$. It is clear that the eigenvalues depend on
$P_{0}$ since the eigenvalue equations are determined for each momentum
subspace connected to $P_{0}$. Similarly, the left eigenvalue equation
among the components $\tilde{\phi}_{P_{0};j}(P_{\nu})$ of the $j$-th
left eigenvector can be written as
\begin{equation}
  \tilde{\phi}_{P_{0};j}(P_{\nu})\{-(K_{\nu+1,\nu}+K_{\nu-1,\nu})\}+
  \tilde{\phi}_{P_{0};j}(P_{\nu-1})K_{\nu,\nu-1}+
  \tilde{\phi}_{P_{0};j}(P_{\nu+1})K_{\nu,\nu+1}=
   z_{P_{0};j}\tilde{\phi}_{P_{0};j}(P_{\nu}).
 \label{eq:left eigenvalue eq}
\end{equation}
\end{widetext}
The left and right eigenvector satisfy following relation,
\begin{equation}
  \tilde{\phi}_{P_{0};j}(P_{\nu})=
  [\varphi_{P_{0}}^{{\rm eq}}(P_{\nu})]^{-1}\phi_{P_{0};j}(P_{\nu}),
  \label{eq:relation between the right and the left eigenvectors}
\end{equation}
where $\varphi_{P_{0}}^{{\rm eq}}(P)$ is given by Eq. (\ref{eq:equilibrium distribution})

The relationship (\ref{eq:relation between the right and the left eigenvectors})
can be easily proved with Eqs. (\ref{eq:right eigenvalue eq}) and (\ref{eq:left eigenvalue eq})
by using the fact that the following detailed balance condition is satisfied:
\begin{equation}
  K_{PP^{\prime}}\cdot\varphi^{{\rm eq}}(P^{\prime})=
   K_{P^{\prime}P}\cdot\varphi^{{\rm eq}}(P).
  \label{eq:detailed balanced condition}
\end{equation}

Keeping the relation (\ref{eq:relation between the right and the left eigenvectors}),
the right eigenvectors $\phi_{P_{0};j}(P_{\nu})$ and the left eigenvectors
$\tilde{\phi}_{P_{0};j}(P_{\nu})$ can be made to satisfy the bi-orthonormality
and bi-completeness relations,
\begin{align}
  \sum_{\nu=-\infty}^{\infty}
   \tilde{\phi}_{P_{0};j}(P_{\nu})\phi_{P_{0};j^{\prime}}(P_{\nu}) &=
  \delta_{j,j^{\prime}}^{{\rm Kr}},\\
  \sum_{\nu=-\infty}^{\infty}
   \tilde{\phi}_{P_{0};j}(P_{\mu})\phi_{P_{0};j}(P_{\nu}) & =
  \delta_{\mu,\nu}^{{\rm Kr}}.
\end{align}
In particular, the right  and left  eigenvectors with zero eigenvalue
are
\begin{equation}
  \phi_{P_{0};0}(P_{\nu})=\varphi_{P_{0}}^{{\rm eq}}(P_{\nu}),
  \label{eq:zero right eigenvector A}
\end{equation}
 and
\begin{equation}
  \tilde{\phi}_{P_{0};0}(P_{\nu})=1,
  \label{eq:zero left eigenvector}
\end{equation}
respectively.
Note that
\begin{equation}
  \sum_{\nu=-\infty}^\infty\phi_{P_{0};0}(P_{\nu})=
  \sum_{\nu=-\infty}^\infty\varphi_{P_{0}}^\mathrm{eq}(P_{\nu})=1,
  \label{eq:normalization of the equilibrium state in each momentum subspace}
\end{equation}
see Eq.\eqref{eq:equilibrium distribution}.

It can be shown that the collision operator (\ref{eq:concrete form of the collision op})
is anti-Hermitian with respect to an inner product weighted by $[\varphi_{P_{0}}^{{\rm eq}}(P_{\nu})]^{-1}$.
For this point see Ref. \cite{2011KKankiSTanakaTPetrosky}. Hence
the eigenvalues of the collision operator (\ref{eq:concrete form of the collision op})
are pure imaginary. Thus, we rewrite the eigenvalues of the collision
operator in Eq. (\ref{eq:k=00003D0eigenvalue problem of the collision operator})
as
\begin{equation}
  z_{P_{0};j}=i\lambda_{P_{0};j},
\end{equation}
 where $\lambda_{P_{0};j}\in\mathbb{R}$.

We solved the eigenvalue problem of the collision operator (\ref{eq:k=00003D0eigenvalue problem of the collision operator})
by numerical diagonalization and continued fraction method. For the
detailed treatments see Ref.   \cite{2010KKankiSTanakaBATayTPetrosky,2011BATayKKankiSTanakaTPetrosky}.

\begin{figure}[htbp]
\begin{raggedright}
\includegraphics[scale=0.35]
{
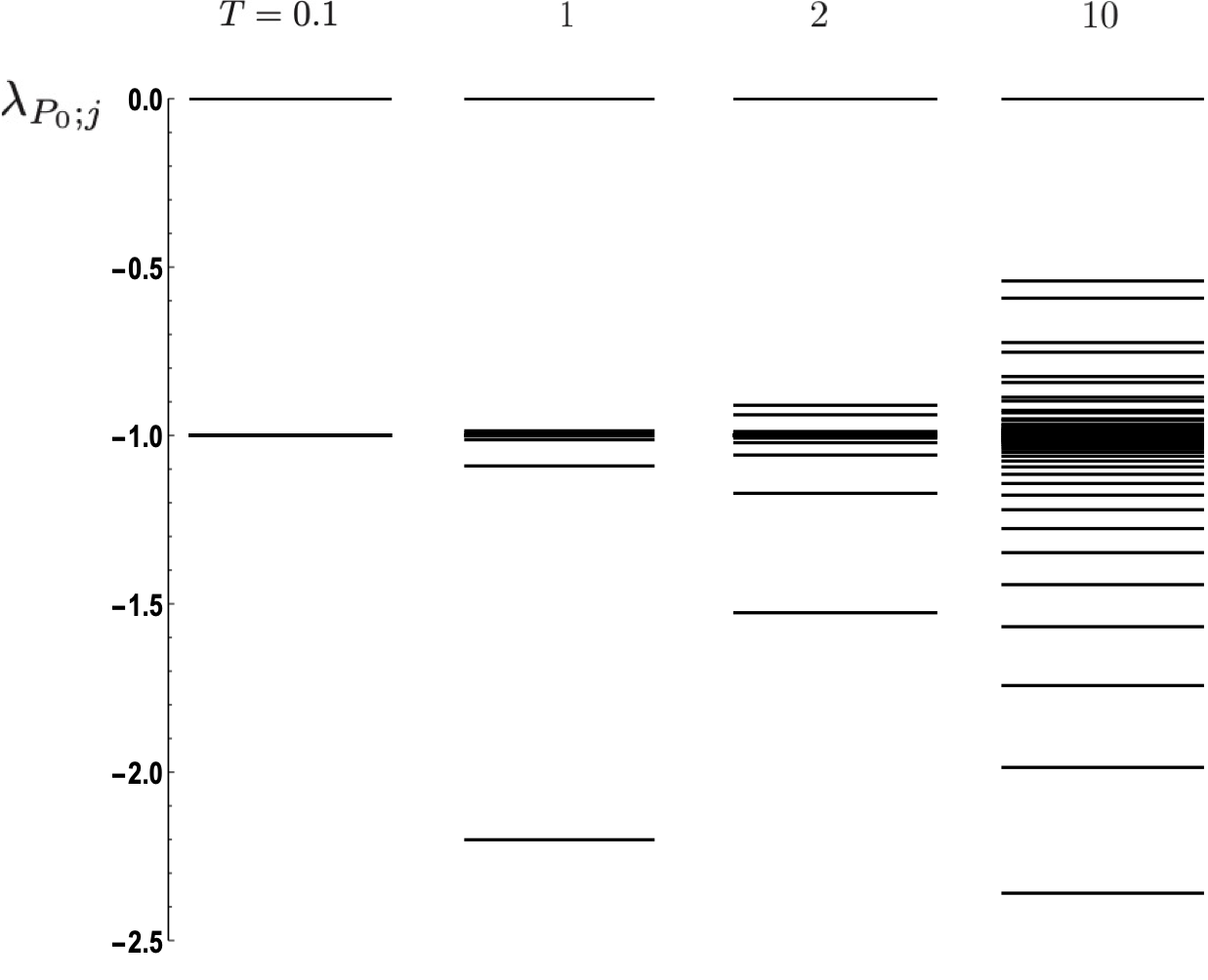}
\par
\end{raggedright}
\caption
 {
 \label{fig:spectrum of K_0 P0=00003D0.5}
 The spectrum of ${\cal K}^{(0)}_P$
 for the temperatures, $T=0.1,1,2,{\rm and},10$,
 and $P_{0}=0.5$ in units where $m=1,c=1,\hbar=1,k_{{\rm B}}=1\ {\rm and}\ |\lambda_{\infty}|=1$.
 }
\end{figure}
\begin{figure}[htbp]
\begin{centering}
\includegraphics[scale=0.6]
{
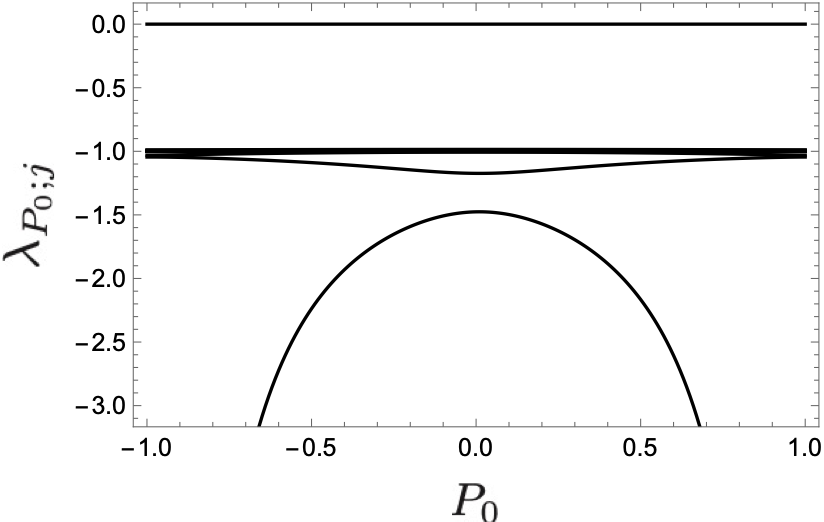}
\caption
 {
 \label{fig:the momentum dependence of the spectrum T=00003D1}
 Momentum dependence of the spectrum of ${\cal K}^{(0)}_P$  when $T=1$.
 The vertical axis is the eigenvalues
 and the horizontal axis is $P_{0}$ in units of $m=1,c=1,\hbar=1,k_{{\rm B}}=1\ {\rm and}\ |\lambda_{\infty}|=1$.
 }
\par
\end{centering}
\end{figure}

The spectrum of ${\cal K}_P^{(0)}$ is obtained for each $P_{0}$ as described
under Eq. (\ref{eq:right eigenvalue eq}). In. Fig. \ref{fig:spectrum of K_0 P0=00003D0.5},
we display the spectrum of ${\cal K}^{(0)}_P$ for several temperatures,
where $P_{0}/mc=0.5$. It is found that the spectrum of ${\cal K}^{(0)}_P$
is discrete and that the spectrum has an accumulation point $\lambda_{\infty}$ given by
\begin{equation}
  \lambda_{\infty}\equiv-\frac{g^{2}m\Delta_{0}^{2}}{\hbar^{2}\rho_{M}c},
\end{equation}
i.e., infinitely many eigenvalues exist in an arbitrarily small neighborhood of $\lambda_{\infty}$.
Thus, we can label the eigenvalues
with all the integers in the following way:
\begin{align}
  \lambda_{P_{0};0} & \equiv
   0>\lambda_{P_{0};1}>\lambda_{P_{0};2}>
   \cdots>\lambda_{P_{0};j}>
   \cdots>\lambda_{\infty}>
   \cdots\nonumber \\& \ \ \ \
   \cdots>\lambda_{P_{0};-j}>
   \cdots>\lambda_{P_{0};-2}>\lambda_{P_{0};-1}>-\infty.
  \label{eq:spectrum of collision op for homogeneous system}
\end{align}
The eigenvalues labeled with $j\ge0$ are the ones which are larger
than $\lambda_{\infty}$, the eigenvalues labeled with $j<0$ are
the ones which are less than $\lambda_{\infty}$, and ${\rm lim}_{j\rightarrow\pm\infty}\lambda_{P_{0};j}=\lambda_{\infty}$.

In Fig. \ref{fig:the momentum dependence of the spectrum T=00003D1},
we display the momentum dependence of the spectrum of ${\cal K}^{(0)}_P$,
where $k_{{\rm B}}T/mc^{2}=1$. The vertical axis is the eigenvalues
measured in units of $|\lambda_{\infty}|$, and the horizontal axis
is $P_{0}/mc$.
There is a zero eigenvalue of the spectrum at each $P_{0}$.
In other words, the collision operators on each momentum subset $P_{0}$ have a collisional invariant.
Therefore, the zero eigenvalues of the
collision operator ${\cal K}_P^{(0)}$ are infinitely degenerate. Moreover,
it is found that the spectrum of the collision operator for the momentum
distribution function has a finite gap between zero eigenvalues and
the non-zero eigenvalues for any momentum $P_{0}$,

\begin{equation}
  \lambda_{{\rm gap}}\equiv
  \lambda_{P_{0};0}-\lambda_{P_{0};1}=-\lambda_{P_{0};1}\neq0.
  \label{lambda_gap}
\end{equation}
 This fact implies that there exists a definite time scale, i.e. the
relaxation time,
\begin{equation}
  \tau_{{\rm rel}}\equiv1/\lambda_{{\rm gap}}={\rm finite},
  \label{eq:relaxation time}
\end{equation}
in momentum relaxation process.
Hence, the local equilibrium situation can be realized in this model.

\hypertarget{B}{}
\section{Derivation of the phenomenological diffusion coefficient with Gaussian
initial condition}

In this section, we derive Eq. (\ref{eq:phenomenological diffusion coefficient in 1D system})
in case where the initial condition is given by a wave function with
the minimum uncertainty.

We rewrite the Wigner distribution function (\ref{eq:explicit form of the Wigner distribution function of the exciton})
as
\begin{equation}
  f^{W}(X,P_{\nu},t)=
  \varphi_{P_{0}}^{{\rm eq}}(P_{\nu})g_{P_{0}}(X,t),
  \label{eq:separated Wigner distribution function}
\end{equation}
 where $\varphi_{P_{0}}^{{\rm eq}}(P_{\nu})$ is defined in Eq. (\ref{eq:equilibrium distribution}),
and
\begin{equation}
  g_{P_{0}}(X,t)\equiv a(P_{0})G(X,P_{0},t),
  \label{eq:product of a weighting factor and spatial distribution}
\end{equation}
 with
\begin{widetext}
\begin{align}
  a(P_{0}) & \equiv
   \sum_{\mu=-\infty}^{\infty}
    \frac{1}{\sqrt{2\pi(\Delta P)^{2}}}
    \exp\left[-\frac{(P_{\mu}(P_{0})-P^{\prime})^{2}}{2(\Delta P)^{2}}\right],
  \label{eq:a(P_{0})}
\end{align}
and
\begin{equation}
  G(X,P_{0},t)\equiv
  \frac{1}{\sqrt{2\pi\{(\Delta X)^{2}+2D(P_{0})t\}}}
  \exp\left[
   -\frac{(X-\sigma(P_{0})t)^{2}}{2\{(\Delta X)^{2}+2D(P_{0})t\}}
  \right].
\end{equation}
\end{widetext}
The function $a(P_{0})$ is a sum of distribution probabilities of the initial momentum distribution within a momentum subspace represented by $P_0$.

Since the momentum state can transition only within each subspace due to the resonance condition in 1D,
the sum of the distribution probabilities within each momentum subspace \eqref{eq:a(P_{0})} is conserved during momentum equilibration.
Therefore,
the distribution along $P$ direction after establishing local equilibrium in phase space
is expressed as the product of the weighting factor and momentum equilibrium distribution: $a(P_{0})\varphi_{P_{0}}^{{\rm eq}}(P_{\nu})$.

On the other hand, $G(X,P_{0},t=0)$
is the initial spatial distribution function given as a Gaussian.

The phenomenological diffusion coefficient (\ref{eq:phenomenological diffusion coefficient})
can be written down in two terms as

\begin{equation}
  D^{(x)}(t)  =
   \frac{1}{2}\frac{d}{dt}
   \left\{ \langle X^{2}\rangle_{t}-\langle X\rangle_{t}^{2}\right\} .
  \label{eq:separated D^(x) (t)}
\end{equation}
 By the definition of the average (See Eq. (\ref{eq:difinition of average 2})),
the first and second term of (\ref{eq:separated D^(x) (t)}) can be
calculated respectively as
\begin{align}
  \langle X^{2}\rangle_{t} & =
   \int_{-mc}^{mc}\!\!\!\!\!\!\!\!dP_{0}
   \sum_{\nu=-\infty}^{\infty}
    a(P_{0})\varphi_{P_{0}}^{{\rm eq}}(P_{\nu})
   \int_{-\infty}^{\infty}\!\!\!\!\!\!\!\!dX\
    X^{2}G(X,P_{0},t),
  \label{eq:first term of D^(x)(t)}\\
  \langle X\rangle_{t}^{2} & =
   \left\{
    \int_{-mc}^{mc}\!\!\!\!\!\!\!\!dP_{0}
    \sum_{\nu=-\infty}^{\infty}
     a(P_{0})\varphi_{P_{0}}^{{\rm eq}}(P_{\nu})
    \int_{-\infty}^{\infty}\!\!\!\!\!\!\!\!dX\
     XG(X,P_{0},t)
   \right\} ^{2}.
  \label{eq:second term of D^(x)(t)}
\end{align}

One can easily integrate over $X$ in both Eq. (\ref{eq:first term of D^(x)(t)})
and Eq. (\ref{eq:second term of D^(x)(t)}) by Gaussian integral, and obtains
\begin{widetext}

\begin{align}
  \langle X^{2}\rangle_{t} & =
   \int_{-mc}^{mc}\!\!\!\!\!\!\!\!dP_{0}
   \sum_{\nu=-\infty}^{\infty}
    a(P_{0})\varphi_{P_{0}}^{{\rm eq}}(P_{\nu})
    \left\{ (\Delta X)^{2}+2D(P_{0})t+(\sigma(P_{0})t)^{2}\right\} ,
  \label{eq:first term after X integration}
\end{align}
\begin{align}
  \langle X\rangle_{t}^{2} & =
   \left\{
    \int_{-mc}^{mc}\!\!\!\!\!\!\!\!dP_{0}
    \sum_{\nu=-\infty}^{\infty}
     a(P_{0})\varphi_{P_{0}}^{{\rm eq}}(P_{\nu})\sigma(P_{0})t
   \right\} ^{2}.
  \label{eq:second term after X integration-1}
\end{align}
 Substituting Eqs. (\ref{eq:first term after X integration}) and
(\ref{eq:second term after X integration-1}) into Eq. (\ref{eq:separated D^(x) (t)}),
we obtain $D^{(x)}(t)$ as a linear function of time:

\begin{align}
  D^{(x)}(t) & =
   \int_{-mc}^{mc}\!\!\!\!\!\!\!\!dP_{0}
   \sum_{\nu=-\infty}^{\infty}
    a(P_{0})\varphi_{P_{0}}^{{\rm eq}}(P_{\nu})D(P_{0})\nonumber \\
 & \ \ \ +
   t\left[
    \int_{-mc}^{mc}\!\!\!\!\!\!\!\!dP_{0}
    \sum_{\nu=-\infty}^{\infty}
     a(P_{0})\varphi_{P_{0}}^{{\rm eq}}(P_{\nu})(\sigma(P_{0}))^{2}-
    \left\{ \int_{-mc}^{mc}\!\!\!\!\!\!\!\!dP_{0}
     \sum_{\nu=-\infty}^{\infty}a(P_{0})
     \varphi_{P_{0}}^{{\rm eq}}(P_{\nu})\sigma(P_{0})
    \right\} ^{2}
    \right].
  \label{eq:D^(x)(t) after X integration}
\end{align}
Since the initial spatial distribution
is given by the Gaussian, the equation
\begin{equation}
  a(P_{0})=\int_{-\infty}^{\infty}\!\!\!\!\!\!\!\!dX\ g_{P_{0}}(X,t),
  \label{eq:normalization condition of spatial distribution}
\end{equation}
 holds. Substituting Eq. (\ref{eq:normalization condition of spatial distribution}) into each terms in Eq. (\ref{eq:D^(x)(t) after X integration}),
 one can obtain the form of the average defined in Eq. (\ref{eq:difinition of average 2}) as
\begin{align}
  D^{(x)}(t) & =
   \int_{-\infty}^{\infty}\!\!\!\!\!\!\!dX
   \int_{-mc}^{mc}\!\!\!\!\!\!\!\!dP_{0}
   \sum_{\nu=-\infty}^{\infty}
    D(P_{0})f^{W}(X,P_{\nu},t) \nonumber\\
  & \ \ \
   +t\left[
   \int_{-\infty}^{\infty}\!\!\!\!\!\!\!dX
   \int_{-mc}^{mc}\!\!\!\!\!\!\!\!dP_{0}
   \sum_{\nu=-\infty}^{\infty}(\sigma(P_{0}))^{2}f^{W}(X,P_{\nu},t)
    -\left\{
     \int_{-\infty}^{\infty}\!\!\!\!\!\!\!dX
     \int_{-mc}^{mc}\!\!\!\!\!\!\!\!dP_{0}
     \sum_{\nu=-\infty}^{\infty}
      \sigma(P_{0})f^{W}(X,P_{\nu},t)
     \right\} ^{2}
   \right].
  \label{eq:D^(x)(t) after X integration 2}
\end{align}
\end{widetext}
Therefore, we get
\begin{equation}
  D^{(x)}(t) =
   \bigl\langle D(P)\bigr\rangle_t+
   t\left[
     \bigl\langle\sigma^{2}(P)\bigr\rangle_t-\bigl\langle\sigma(P)\bigr\rangle_t^{2}
   \right].
  \label{eq:Gaussian initial condition phenomenological diffusion coefficient}
\end{equation}
Then, we can integrate over $X$ in the averages in Eq. (\ref{eq:Gaussian initial condition phenomenological diffusion coefficient}) and obtain
\begin{align}
  D^{(x)}(t)&=
   \bigl\langle D(P)\bigr\rangle_{\rm eq}+
   t\left[
     \bigl\langle\sigma^{2}(P)\bigr\rangle_{\rm eq}-\bigl\langle\sigma(P)\bigr\rangle_{\rm eq}^{2}
   \right]
\nonumber\\
&=
\bar{D}+t\left\langle (\sigma(P)-\bar{\sigma})^{2}\right\rangle_{\rm eq},
\end{align}
where
\begin{align}
  \bar{D}&=
  \bigl\langle D(P)\bigr\rangle_{\rm eq}\equiv
  \int_{-mc}^{mc}\!\!\!\!\!\!\!\!dP_{0}
  \sum_{\nu=-\infty}^{\infty}
   a(P_{0})\varphi_{P_{0}}^{{\rm eq}}(P_{\nu})D(P_{0}),
  \label{eq:<D(P)>_ell}\\
%
%
  \bar{\sigma}&=
  \bigl\langle\sigma(P)\bigr\rangle_{\rm eq}\equiv
  \int_{-mc}^{mc}\!\!\!\!\!\!\!\!dP_{0}
   \sum_{\nu=-\infty}^{\infty}a(P_{0})
   \varphi_{P_{0}}^{{\rm eq}}(P_{\nu})\sigma(P_{0}).
  \label{eq:<sigma(P)>_ell}
\end{align}

Let us prove that Eqs. (\ref{eq:<>_ell=<>_0})
is satisfied.
Using the relation (\ref{eq:normalization of the equilibrium state in each momentum subspace}),
Eq. (\ref{eq:<D(P)>_ell}) can be reduced
as
\begin{align}
  \bigl\langle D(P)\bigr\rangle_{\rm eq} & =
   \int_{-mc}^{mc}\!\!\!\!\!\!\!\!dP_{0}\
    a(P_{0})D(P_{0}).
  \label{eq:average over momentum distribution at local eq 2}
\end{align}
Substituting Eq. (\ref{eq:a(P_{0})})
into Eq. (\ref{eq:average over momentum distribution at local eq 2}),
we get
\begin{align}
  \bigl\langle D(P)\bigr\rangle_{\rm eq}=&\nonumber\\
   \int_{-mc}^{mc}\!\!\!\!\!\!\!\!dP_{0}
   \sum_{\mu=-\infty}^{\infty}&
    \frac{1}{\sqrt{2\pi(\Delta P)^{2}}}\exp\left[-\frac{(P_{\mu}(P_{0})-P^{\prime})^{2}}{2(\Delta P)^{2}}\right]D(P_{0}).
\end{align}
Replacing the integration over $P_{0}$ and the summation over all the discrete momentum $P_{\mu}$ with an integration over $P$, we obtain

\begin{align}
  \bigl\langle D(P)\bigr\rangle_{\rm eq} & =
  \int_{-\infty}^{\infty}\!\!\!\!\!\!\!\!dP\
   \frac{1}{\sqrt{2\pi(\Delta P)^{2}}}
   \exp\left[-\frac{(P-P^{\prime})^{2}}{2(\Delta P)^{2}}\right]D(P)
  \label{eq:barDT}\\
  &=
  \bigl\langle D(P)\bigr\rangle_{t=0}\nonumber.
\end{align}

 Similarly, one can get the averages of the hydrodynamic sound velocity as
\begin{align}
  \bigl\langle\sigma(P)\bigr\rangle_{\rm eq}&=
  \int_{-\infty}^{\infty}\!\!\!\!\!\!\!\!dP\
   \frac{1}{\sqrt{2\pi(\Delta P)^{2}}}
   \exp\left[-\frac{(P-P^{\prime})^{2}}{2(\Delta P)^{2}}\right]\sigma(P)
  \label{eq:barsigmaT}\\
  &=
  \bigl\langle\sigma(P)\bigr\rangle_{t=0}\nonumber.
\end{align}

\hypertarget{C}{}
\section{phenomenological diffusion coefficient in arbitrary initial condition}

In this section, we show that the phenomenological diffusion coefficient
$D^{(x)}(t)$ defined as Eq. (\ref{eq:phenomenological diffusion coefficient})
increases linearly with time in a case where the initial condition
is given as arbitrary square integrable function.
To prove this we
use a theorem shown by Calcaterra (\cite{2008CCalcaterra,2008CCalcaterraABoldt},
Theorem 1). Here, we introduce the theorem.

$\textit{Definition}s$: $L^{2}(\mathbb{R})$ $denotes$ $the$ $space$
$of$ $square$ $integrable$ $functions$ $f$: $\mathbb{R}\rightarrow\mathbb{R}$
$with$ $norm$
\begin{equation}
  ||f||_{2}\equiv\sqrt{\int_{\mathbb{R}}|f(x)|^{2}dx}.\nonumber
\end{equation}
$Relation$ $f\underset{\epsilon}{\approx}g$ $means$ $||f-g||_{2}<\epsilon$.

$\textit{Theorem 1}$: $For$ $any$ $f\in L^{2}(\mathbb{R})$ $and$
$any$ $\epsilon>0$ $there$ $exists$ $s>0$ , $N\in\mathbb{N}$
$and$ $a_{n}\in\mathbb{R}$ $such$ $that$
\[
f\underset{\epsilon}{\approx}\sum_{n=0}^{N}a_{n}e^{-(x-ns)^{2}}.
\]

Calcaterra gives one choice of coefficients as
\begin{equation}
  a_{n}
   =\frac{1}{\sqrt{\pi}}\frac{(-1)^{n}}{n!}
   \sum_{k=n}^{N}
    \frac{1}{(k-n)!(2s)^{k}} \int_{\mathbb{R}}f(x)e^{x^{2}}\frac{d^{k}}{dx^{k}}e^{-x^{2}}dx.
\end{equation}

We obtained the formal solution of Fourier component of the Wigner
function belonging to a certain momentum subspace $P_{0}$ for $t\gtrsim\tau_{{\rm rel}}$
as (see Eq. (\ref{eq:distribution fnc after local eq}) 
)

\begin{widetext}
\begin{equation}
  f_{k}(P_{\nu},t\gtrsim\tau_{{\rm rel}})=
   e^{-iz_{P_{0};0}^{(k)}t}\phi_{P_{0};0}^{(k)}(P_{\nu})
   \sum_{\mu=-\infty}^{\infty}
    \tilde{\phi}_{P_{0};0}^{(k)}(P_{\mu})f_{k}(P_{\mu},t=0).
  \label{eq:Appendix distribution fnc after local eq in the hydrodynamic regime}
\end{equation}
 We can approximate Eq. (\ref{eq:Appendix distribution fnc after local eq in the hydrodynamic regime})
in the hydrodynamic regime as mentioned at Eqs. \eqref{eq:distribution fnc after local eq in the hydrodynamic regime}
and \eqref{eq:explicit Fourier component at local eq},
\begin{align}
  f_{k}(P_{\nu},t\gtrsim\tau_{{\rm rel}}) & \simeq
   e^{-ik\sigma(P_{0})t}e^{-k^{2}D(P_{0})t}\varphi_{P_{0}}^{{\rm eq}}(P_{\nu})
   \sum_{\mu=-\infty}^{\infty}f_{k}(P_{\mu},t=0).
  \label{eq:fk distribution in arbitrary initial condition-1}
\end{align}

Equation (\ref{eq:fk distribution in arbitrary initial condition-1})
is the formal solution of Fourier component of the Wigner distribution function
after establishing local equilibrium in the hydrodynamic regime.
We define the
factor dependent on the initial distribution as
\begin{equation}
  W_{P_{0}}(k)\equiv\sum_{\mu=-\infty}^{\infty}f_{k}(P_{\mu}(P_{0}),t=0).
\end{equation}
\end{widetext}
The factor $W_{P_{0}}(k)$ is
the function defined for each value of $P_0$.

By performing a Fourier transform on Eq. (\ref{eq:fk distribution in arbitrary initial condition-1}),
we obtain the formal solution of the Wigner distribution function as
\begin{equation}
  f^{W}(X,P_{\nu},t\gtrsim\tau_{{\rm rel}})=
   \varphi_{P_{0}}^{{\rm eq}}(P_{\nu})g_{P_{0}}(X,t),
  \label{eq:the formal solution of the Wigner distribution function}
\end{equation}
 where
\begin{equation}
  g_{P_{0}}(X,t)\equiv
   \frac{1}{2\pi}\int_{-\infty}^{\infty}\!\!\!\!\!dk\
    e^{ikX}e^{-ik\sigma(P_{0})t}e^{-k^{2}D(P_{0})t}W_{P_{0}}(k).
  \label{eq:space distribution of Wigner distribution}
\end{equation}

Equation (\ref{eq:the formal solution of the Wigner distribution function}) shows
that the Wigner distribution function in a certain momentum subspace $P_{0}$ can
be written in the form of the product of the momentum equilibrium distribution
$\varphi_{P_{0}}^{{\rm eq}}(P_{\nu})$ and the
spatial distribution (\ref{eq:space distribution of Wigner distribution}).

Putting $t=0$ in Eq. (\ref{eq:space distribution of Wigner distribution}),
we get the initial spatial distribution function defined for each value of $P_0$
as
\begin{equation}
  g_{P_{0}}(X,0)=
  \frac{1}{2\pi}\int_{-\infty}^{\infty}\!\!\!\!\!dk\ e^{ikX}W_{P_{0}}(k).
  \label{eq:relation between gx and Wk}
\end{equation}
 Here, we use Theorem 1. According to Theorem 1, any square integrable
function can be approximated with arbitrary precision by the linear
combination of Gaussians with a single variance. If we assume $g_{P_{0}}(X,0)$
as square integrable function, it can be expanded as
{[}cf. Eq. (\ref{eq:product of a weighting factor and spatial distribution}){]}
\begin{equation}
  g_{P_{0}}(X,0)\underset{\epsilon}{\approx}
   \sum_{n=0}^{N}
    \frac{1}{\sqrt{2\pi}\Delta X}
    a_{n}(P_{0})e^{-(\frac{X}{\sqrt{2}\Delta X}-ns)^{2}},
  \label{eq:expansion of the initial spatial distribution function}
\end{equation}
 where $\Delta X$ is a constant with a unit of length, and
\begin{widetext}
\begin{equation}
  a_{n}(P_{0})\equiv
   \frac{(-1)^{n}}{n!}
   \sum_{k=n}^{N}\frac{1}{(k-n)!(2s)^{k}}
   \int_{-\infty}^{\infty}g_{P_{0}}(X,0)e^{X^{2}}
    \frac{d^{k}}{dX^{k}}e^{-X^{2}}dX.
  \label{eq:one choice of coefficients}
\end{equation}

By inverse transformation of Eq. (\ref{eq:relation between gx and Wk}),
we obtain
\begin{equation}
  W_{P_{0}}(k)=
  \sum_{n=0}^{N}
   a_{n}(P_{0})\exp\left[-\frac{(\Delta X)^{2}}{2}k^{2}-i\sqrt{2}\Delta Xnsk\right].
  \label{eq:weighting factor expanded by Gaussian}
\end{equation}
 Substituting Eq. (\ref{eq:weighting factor expanded by Gaussian})
into Eq. (\ref{eq:space distribution of Wigner distribution}) and
integrating over $k$, we get spatial distribution function in a certain
momentum subspace $P_{0}$ as

\begin{equation}
  g_{P_{0}}(X,t)=
   \sum_{n=0}^{N}
    a_{n}(P_{0})\frac{1}{\sqrt{2\pi((\Delta X)^{2}+2D(P_{0})t)}}
    \exp\left[-\frac{(X-\sigma(P_{0})t-\sqrt{2}\Delta Xns)^{2}}{2\{(\Delta X)^{2}+2D(P_{0})\}t}\right].
  \label{eq:the time evolution function of the spatial distribution function}
\end{equation}
We note that $g_{P_{0}}(X,t)$ is represented by a single term in the case where
the initial distribution is given by one Gaussian as shown in Eq. (\ref{eq:product of a weighting factor and spatial distribution}).

\end{widetext}

Now we can calculate
the phenomenological diffusion coefficient (\ref{eq:phenomenological diffusion coefficient})
with 
Eqs. (\ref{eq:the formal solution of the Wigner distribution function}) and
(\ref{eq:the time evolution function of the spatial distribution function}).
Since we got the spatial distribution $g_{P_{0}}(X,t)$ as the linear
combination of Gaussians, we can integrate over $X$ by Gaussian integral.
After simple calculation, the phenomenological diffusion coefficient
$D^{(x)}(t)$ in a general case is finally expressed as
\begin{align}
  D^{(x)}(t) =&
   \bar{D}+t\ \bigl\langle (\sigma(P)-\bar{\sigma})^{2}\bigr\rangle_{\rm eq}
   \nonumber\\&
   +
   \bigl\langle
    \left(X-\langle X\rangle_{t=0})
         (\sigma(P)-\bar{\sigma}\right)
   \bigr\rangle _{t=0},
 \label{eq:General phenomenological diffusion coefficient}
\end{align}
where the notation $\langle\cdots\rangle_{t=0}$ indicates to take average over the initial Wigner distribution function. Note that
\begin{equation}
  \langle X\rangle_{t=0}\neq
  \langle X\rangle_{\rm eq}.
\end{equation}
The averages of transport coefficients in Eq. (\ref{eq:General phenomenological diffusion coefficient})
are expressed as {[}cf. Eq. (\ref{eq:average over momentum distribution at local eq 2}){]}
\begin{equation}
  \bar{D}=\bigl\langle D(P)\bigr\rangle_{\rm eq}=
  \sum_{n=0}^{N}\int_{-mc}^{mc}\!\!\!\!\!\!\!dP_{0}\ a_{n}(P_{0})D(P_{0}),
 \label{eq:general D(P) average}
\end{equation}
\begin{equation}
  \bar{\sigma}=\bigl\langle\sigma(P)\bigr\rangle_{\rm eq}=
  \sum_{n=0}^{N}\int_{-mc}^{mc}\!\!\!\!\!\!\!dP_{0}\ a_{n}(P_{0})\sigma(P_{0}).
  \label{eq:general sigma(P) average}
\end{equation}

Equation (\ref{eq:General phenomenological diffusion coefficient})
shows that $D^{(x)}(t)$ increases linearly with time and diverges
in the long-time limit. The third term in the right-hand side of Eq.
(\ref{eq:General phenomenological diffusion coefficient}) comes from
the deviation of the initial distribution from Gaussian distribution.
The third term vanishes in cases where the initial spatial distribution
is given by one Gaussian distribution since the term is an integral
of an odd function of $X$ in those cases. Under that condition, Eq.
(\ref{eq:General phenomenological diffusion coefficient}) therefore
reduces to Eq. (\ref{eq:Gaussian initial condition phenomenological diffusion coefficient}).



\end{document}